\documentclass[aps,prb,reprint]{revtex4-1}%
\usepackage{amsmath}
\usepackage{array}
\usepackage{array}
\usepackage{amsfonts}
\usepackage{amssymb}
\usepackage{graphicx}
\usepackage{color}
\usepackage{dcolumn}
\usepackage{bm}
\usepackage{hyperref}
\usepackage{epstopdf}%
\providecommand{\U}[1]{\protect\rule{.1in}{.1in}}
\setcitestyle{super}
\newcommand*{\citen}
[1]{\begingroup\romannumeral-`\x
\setcitestyle{numbers}
[\cite{#1}]\endgroup
}

\begin{document}

\title{Crystal field effects on spin pumping}
\author{Adam B. Cahaya$^{1}$, Alejandro~O. Leon$^{1}$, and Gerrit E.~W. Bauer$^{2,3}$}
\date{\today }

\affiliation{$^1$Institute for Materials Research, Tohoku University, Sendai
980-8577, Japan} 
\affiliation{$^2$WPI-AIMR $\&$ CSRN, Tohoku University, Sendai
980-8577, Japan} 
\affiliation{$^3$Zernike Institute for Advanced Materials,
Groningen University, The Netherlands}

\begin{abstract}
\textquotedblleft Spin pumping\textquotedblright\ is the injection of spin
angular momentum by a time-dependent magnetization into an adjacent normal
metal proportional to the spin mixing conductance. We study the role of
electrostatic interactions in the form of crystal fields on the pumped spin
currents generated by insulators with exchange-coupled local moments at the
interface to a metal. The crystal field is shown to render the spin currents
anisotropic, which implies that the spin mixing conductance of insulator $|$
normal metal bilayers depends on crystal cut and orientation. We interpret the
interface \textquotedblleft effective field\textquotedblright\ (imaginary part
of the spin mixing conductance) in terms of the coherent motion of the
equilibrium spin density induced by proximity in the normal metal.
\end{abstract}
\maketitle

\section{Introduction}
The interaction between the magnetization and currents in small structures and
devices has attracted much attention in the last two decades. The generation
of a spin current by magnetization dynamics is referred to as spin
pumping~\citen{Tserkovnyak,Simanek}: a time-dependent magnetization
\textquotedblleft pumps\textquotedblright\ a spin current with magnitude and
polarization $\mathbf{J}=g_{r}^{\uparrow\downarrow}\mathbf{m}\times
\dot{\mathbf{m}}-g_{i}^{\uparrow\downarrow}\dot{\mathbf{m}}$ into a normal
metal contact, where $\mathbf{m}$ is the unit magnetization vector,
$\mathbf{\dot{m}}$ its time derivative, and $g^{\uparrow\downarrow}%
=g_{r}^{\uparrow\downarrow}+ig_{i}^{\uparrow\downarrow}$ is the complex
interfacial spin mixing conductance. The spin pumping enhances the
magnetization damping, and can be interpreted as the Onsager reciprocal effect
to the current-induced spin transfer torque, both being governed by the same
spin mixing conductance~\citen{Brataas2012}. The mixing conductance of the
magnetic insulator yttrium iron garnet (YIG) was predicted to be of the same
order of magnitude as that of magnetic metals~\citen{Jia}, which was
subsequently confirmed by experiments \citen{Burrowes,Weiler}. A dependence of
the spin mixing conductance on the interface cut and orientation to the normal
metal has also been predicted~\citen{Jia} and
confirmed~\citen{Casanova,Tokac}. This anisotropy could partly be explained by
the density of the local Fe magnetic moment directly at the interface. The
rotational symmetry of magnetic atoms can be broken by the electric fields
generated by neighboring atoms, i.e. the so called \textit{crystal field}. The
relationship between the spin pumping and the local symmetry of magnetic
moments at the interface has, to the best of our knowledge, not been studied
yet. We therefore focus here on noncubic crystal fields of 3d transition
metal ions with partially (not fully or half-) filled shells. These are
predicted to cause effects that are much stronger than those generated by a
cubic crystal field or when acting on 4f moments. Also, in the former case the
spin orbit interaction is much weaker than the spin orbit interaction that we,
hence, disregard here.

Under crystal fields, the angular part of the single 3d electron is described
by the real valued doubly degenerate $e_{g}$ and triply degenerate $t_{2g}$
orbitals~\citen{BookMagnetismBlundell,BookMagnetismTMO}. For transition metal
ions on sites with octahedral symmetry, the energy level order is $E_{e_{g}%
}>E_{t_{2g}}$ while in tetrahedral environment $E_{t_{2g}}>E_{e_{g}}%
$~\citen{BookMagnetismBlundell,BookMagnetismTMO}. The total orbital angular
momentum in this basis is quenched, $\left\langle L_{z}\right\rangle
=0$~\citen{BookMagnetismBlundell,BookMagnetismTMO}. The magnetism is then
predominantly caused by the electron (Pauli) spins. {When the spin orbit
interaction is not negligible but competes with the crystal fields, the
eigenstates are complex combinations of the sets $e_{g}$ and $t_{2g}$. The
orbital moment is then not completely quenched $\left\langle L_{z}%
\right\rangle \neq0$ and the energy depends on the direction of the
magnetization relative to the crystal axes (magnetic anisotropy)
\citen{BookSkomski1}. The effects of the spin-orbit interaction is discussed
in a forthcoming paper with emphasis on partially filled 4f shells
\citen{Ale}.}

Here we study the role of crystal fields on the spin mixing conductance that
governs spin pumping and other properties of interfaces. {The paper is
organized as follows. In Sec.~\ref{Sec:singleion}, we review the static and
dynamics of 3d transition metal magnetic moments, disregarding their weak
spin-orbit interactions. In the presence of crystal fields, the ground state
electronic density of individual ions is nonspherical. By the exchange
interaction such local moment induces in a metal an oscillating proximity spin
density and associated Ruderman-Kittel-Kasuya-Yosida interaction (RKKY)
\citen{RudermanKittel,
Kasuya, Yosida} that are also anisotropic. This has, for example, been
confirmed by first principles calculations on metallic surfaces \citen{Zhou}.
The effects of such anisotropies on the spin dynamics are discussed in
Sec.~\ref{Sec:spincurrent}, where we find that the spin current emitted by a
dynamic magnetization is enhanced in certain directions. We discuss how the
anisotropy influences local magnetization dynamic in term of enhanced damping
in Sec.~\ref{Sec:damping}. In Sec.~\ref{Sec:interface}, we extend our analysis
to magnetic insulators in which the local moments at the interface are exposed
to normal metal contacts. In Sec. \ref{SecConclusions} we conclude that the
crystal fields induce differences in the spin pumping for different crystal
growth directions, which might help to explain some experiments. }
\section{Single-ion model}

\label{Sec:singleion} Consider a single localized magnetic moment generated by
a partially filled 3d shell with spin density $\mathbf{S}_{d}(\mathbf{r},t)$
that depends adiabatically on time. In terms of the single electron wave
functions $\psi_{j}(\mathbf{r})$ with orbital index $j$, the ground state spin
density reads
\begin{equation}
\mathbf{S}_{d}(\mathbf{r},t)=\mathbf{S}(t)n_{d}(\mathbf{r}), \label{Sdrt}%
\end{equation}
where the electron density distribution of unpaired electrons
\begin{align}
n_{d}(\mathbf{r})  &  =\sum_{j}{S}_{j}|\psi_{j}(\mathbf{r})|^{2},
\end{align}\begin{align}
S_{j}  &  =\frac{f_{j,\uparrow}-f_{j,\downarrow}}{\sum_{k}\left(
f_{k,\uparrow}-f_{k,\downarrow}\right)  } \label{Socc}%
\end{align}
depends on the occupation numbers $f_{j,m_{s}}$ of orbital $j$ and spin label
$m_{s}=\left\{  \uparrow,\downarrow\right\}  $\ and is normalized, $\int
d\mathbf{r}\ n_{d}(\mathbf{r})=1.$ The occupation numbers $f_{j,m_{s}}$ are
governed by the \textit{aufbau} principle when the thermal energy is much
smaller than the crystal field splitting ($\Delta$), ie. $k_{B}T/\left\vert
\Delta\right\vert \ll1$, where $k_{B}$ is the Boltzmann constant and $T$ the
temperature. Since spin orbit interaction is disregarded, the time dependence
is encoded exclusively in the unit vector of the total spin $\mathbf{S}(t).$

\begin{table}[b]
\centering
\caption{The deformation of the spin density of 3d orbitals can be expressed
in terms of the quadrupole moment $Q_{2}$, which is obtained from this table
and the occupation numbers.}\label{Table3d1}%
\begin{tabular}
[c]{lr}\hline
\hline
& \\[-2ex]%
\text{Orbital} $Y_{j}$ & $\left\langle \Omega\right\rangle _{j}$\\[1ex]\hline
& \\[-2ex]%
$Y_{z^{2}}$ & $4/7$\\[1ex]
& \\[-2ex]%
$Y_{x^{2}-y^{2}}$ and $Y_{xy}$ & $-4/7$\\[1ex]
& \\[-2ex]%
$Y_{xz}$ and $Y_{yz}$ & $2/7$\\[1ex]\hline\hline
\end{tabular}
\end{table}

In the presence of crystal fields, $n_{d}(\mathbf{r})$ has the point symmetry
of the crystal site (or higher) that is characterized by a multipolar
expansion. Here we focus on the common case of uniaxial deformation along the
$z$ direction, which allows parameterizing of the anisotropy in the spin
density by its quadrupole moment%
\begin{equation}
Q_{2}=\int d\mathbf{r}r^{2}\left(  \frac{3z^{2}}{r^{2}}-1\right)
n_{d}(\mathbf{r}), \label{quadrupole}%
\end{equation}
where $z=r\cos\theta$ is the coordinate along the symmetry axis of
$n_{d}(\mathbf{r})$. $Q_{2}>0\left(  <0\right)  $ describes a prolate (oblate)
ellipsoid-like distribution. Decomposing the orbitals in the radial and
angular functions, $\psi_{j}(\mathbf{r})=R_{3d}(r)Y_{j}(\boldsymbol{\Omega})$,
the quadrupole reads
\begin{equation}
Q_{2}=\langle r^{2}\rangle\sum_{j}S_{j}\int\left(  3\cos^{2}\theta-1\right)
|Y_{j}(\mathbf{\Omega})|^{2}d\boldsymbol{\Omega}, \label{quadrupolev2}%
\end{equation}
where $r=\left\vert \mathbf{r}\right\vert $, $\langle r^{2}\rangle\equiv\int
r^{2}drr^{2}R_{3d}(r)$, $\mathbf{\Omega}\equiv\mathbf{r}/r$ and
$d\boldsymbol{\Omega}=d\theta d\phi\sin\theta$. The radial function
$R_{3d}(r)$ can be approximated by Slater-type orbitals~\citen{freeman,waber},
while the angular function are linear combinations of spherical harmonics (see
Appendix~\ref{AppendixYj}). $Q_{2}$ is calculated using the occupation numbers
and the coefficients $\left\langle \Omega\right\rangle _{j}=\int\left(
3\cos^{2}\theta-1\right)  |Y_{j}(\boldsymbol{\Omega})|^{2}d\boldsymbol{\Omega
}$, listed in Table~\ref{Table3d1}. 

\begin{figure}[tbh!]
\centering
\includegraphics[width=\columnwidth]{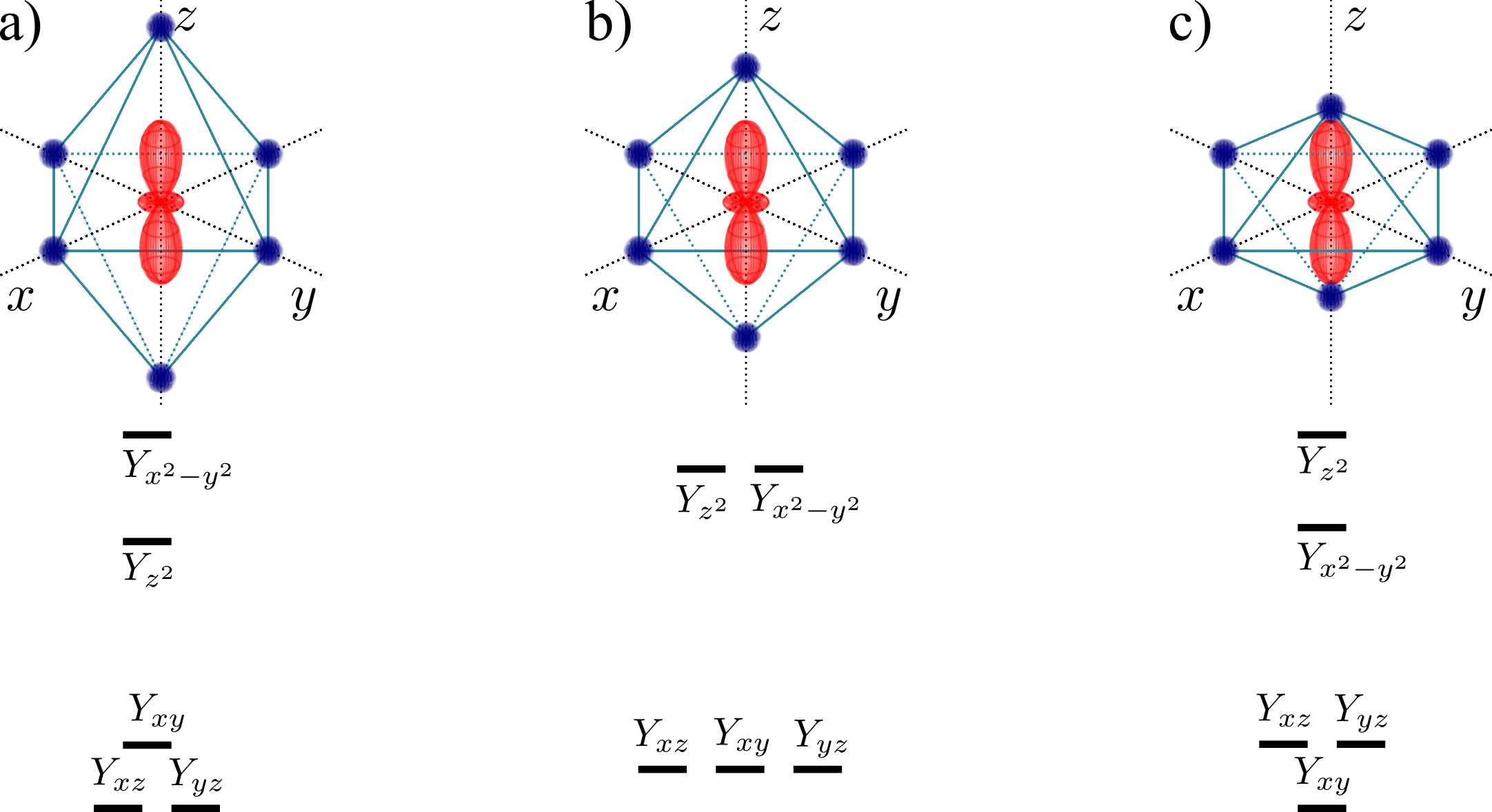}\caption{(Color online)
The{\ $Y_{z^{2}}$ orbital of a 3d magnetic ion in an octahedral environment
(upper panel) and the corresponding orbital splitting of the real-valued
orbitals $e_{g}$($Y_{z^{2}}$, $Y_{x^{2}-y^{2}}$) and $t_{2g}$($Y_{xy}$,
$Y_{xz}$, $Y_{yz}$)} (lower panel). The octahedral environment is a) elongated
b) unperturbed and c) compressed in the $z$ direction. }%
\label{figOct}%
\end{figure}

Crystal fields can be parameterized by a \emph{point charge model} of the
local environment. The Hamiltonian close to the center of an octahedron made
from point charges $qe$ {is
\begin{align}
H_{cf}(\mathbf{r})  &  =\sum_{n}\frac{-qe^{2}}{4\pi\varepsilon_{0}\left\vert
\mathbf{r}-\mathbf{r}_{n}\right\vert }\simeq\frac{-3qe^{2}}{2\pi
\varepsilon_{0}R_{0}}+\nonumber\\
&  \Delta_{\mathrm{octa}}\frac{x^{4}+y^{4}+z^{4}-3(x^{2}y^{2}+x^{2}z^{2}%
+y^{2}z^{2})}{\langle r^{4}\rangle}, \label{Eq_cfpoints}%
\end{align}
where the subscript $n$ labels the point charges at{ $\{(\pm R_{0},0,0),(0,\pm
R_{0},0),(0,0,\pm R_{0})\}$}, $\mathbf{r}=x\mathbf{\hat{\mathbf{x}}%
}+y\mathbf{\hat{y}}+z\mathbf{\hat{z}}$, and the Cartesian axes $\{x,y,z\}$ are
oriented along the crystallographic directions (100), (010) and (001),
respectively. $R_{0}$ is the nearest-neighbor distance, $\varepsilon_{0}$ is
the vacuum permittivity, ${-}${$e\ $is the electron charge, and }$qe$ is the
electric charge of neighboring ions. In metals, ion cores are positively
charged, i.e. $q>0$, while in transition metal oxides the oxygen anions
dominate and $q<0$. The crystal field parameter is $\Delta_{\mathrm{octa}%
}=-{7qe^{2}\langle r^{4}\rangle R_{0}^{-5}}\left(  {8\pi\varepsilon_{0}%
}\right)  ^{-1}$ and can be estimated as $\Delta\sim2$ eV for $q=-2$, $\langle
r^{4}\rangle^{1/4}=1.5$ \AA , $R_{0}=3$ \AA . In the tetrahedral site, on the
other hand, the magnetic atom sits in the center of a cube defined by $(\pm
R_{0},\pm R_{0},\pm R_{0})/\sqrt{3}$. Both octahedral and tetrahedral sites
are described by the same Hamiltonian~\citen{BookMagnetismTMO} but
$\Delta_{\mathrm{tetra}}=-4\Delta_{\mathrm{octa}}/9$.} {Figure~\ref{figOct}b)
shows the crystal field splitting for a symmetric octahedron (charges
equidistant from the origin)} into a doubly degenerate $e_{g}$ and triply
degenerate $t_{2g}$ states (see Appendix). Uniaxial strain breaks the
degeneracies of $e_{g}$ and $t_{2g}$ levels
\citen{JahnTeller,Auxetic,Auxetic2} as sketched in Figure~\ref{figOct}a) and c).

Half-filled shells, such as Mn$^{2+}$ and Fe$^{3+}$are isotropic (spherical)
and their $Q_{2}$ vanishes in any crystal field. The quadrupolar moments
vanish as well for octahedral and tetrahedral crystal fields, because the
half-filled $e_{g}$ and $t_{2g}$ shells are still nearly spherical:
\begin{align}
e_{g}  &  :\int d\boldsymbol{\Omega}\left(  3\cos^{2}\theta-1\right)  \left(
|Y_{z^{2}}|^{2}+|Y_{x^{2}-y^{2}}|^{2}\right)  =0,\\
t_{2g}  &  :\int d\boldsymbol{\Omega}\left(  3\cos^{2}\theta-1\right)  \left(
|Y_{xy}|^{2}+|Y_{xz}|^{2}+|Y_{yz}|^{2}\right)  =0.
\end{align}
{\ The quadrupole in the presence of compressive and tensile uniaxial strains
depends on the occupation numbers as
\begin{equation}
\frac{Q_{2}}{\left\langle r^{2}\right\rangle }=\frac{2}{7}\left[  S_{Y_{xz}%
}+S_{Y_{yz}}+2\left(  S_{Y_{z^{2}}}-S_{Y_{x^{2}-y^{2}}}-S_{Y_{xy}}\right)
\right]  , \label{Q2asoccupation}%
\end{equation}
where $S_{Y_{j}}$ is given by Eq.~(\ref{Socc}).} We note that even in
distorted octahedral sites, some ions such as V$^{2+}$, Cr$^{3+}$ Ni$^{2+}$
and Cu$^{3+}$ have $Q_{2}=0$ because the $e_{g}$ and $t_{2g}$ are half filled.

\subsection*{Interaction between a magnetic ion and conduction electrons}

The interaction between localized magnetic moments and conduction electrons
with spin density $\mathbf{s}_{c}(\mathbf{r},t)$ is described by the s-d
exchange Hamiltonian~\citen{Larsen, Davis}. In the local-density
approximation:
\begin{equation}
H_{\mathrm{s-d}}=-\frac{J}{\hbar^{2}}\int\mathbf{S}_{d}(\mathbf{r}%
,t)\cdot\mathbf{s}_{c}(\mathbf{r},t)d\mathbf{r}, \label{Hsd}%
\end{equation}
where in the static and strong screening limit of the Coulomb interaction the
exchange constant $J=g_{e}^{-1}$ is the reciprocal conduction electron density
of states $g_{e}$ of the host metal and $\hbar$ is Planck's constant divided
by $2\pi$. For free electrons, $g_{e}=m_{e}k_{F}/(\pi^{2}\hbar^{2}%
)=3n_{e}/(2E_{F})$, in terms of the electron density $n_{e}$, Fermi energy
$E_{F}$ and the effective electron mass $m_{e}$. In the ground state, a static
magnetic moment induces spin density oscillations. $H_{\mathrm{s-d}}$ also
communicates the time-dependence of the magnetic moment $\mathbf{\dot{S}}%
\neq0$ to the conduction electrons, which can be formulated by extending the
RKKY perturbation theory into the time domain~\citen{Simanek}. Magnetization
dynamics can be excited by magnetic or spin resonance, but also by spin
transfer torques due to voltage and temperature gradients, lattice vibrations, etc.~\citen{Adachi,Cahaya}.

For sufficiently weak coupling, the response of the conduction electrons to a
time dependent local moment $\mathbf{S}_{d}(\mathbf{r},t)$ reads
\begin{equation}
\mathbf{s}_{c}(\mathbf{r},t)=\frac{J}{\hbar^{2}}\int d\mathbf{r}^{\prime
}dt^{\prime}\chi\left(  \mathbf{r}-\mathbf{r}^{\prime},t-t^{\prime}\right)
\mathbf{S}_{d}(\mathbf{r}^{\prime},t^{\prime}),
\end{equation}
where $\chi\left(  \mathbf{r},t\right)  $ is the (scalar) dynamic spin
susceptibility of the homogeneous host metal. In frequency and momentum space%
\begin{equation}
\mathbf{s}_{c}(\mathbf{q},\omega)=\frac{J}{\hbar^{2}}\chi\left(
q,\omega\right)  \mathbf{S}_{d}\left(  \mathbf{q},\omega\right)  ,
\label{EqScSingleEq}%
\end{equation}
where
\begin{align}
f\left(  \mathbf{q},\omega\right)   &  =\int d\mathbf{r}\int dtf\left(
\mathbf{r},t\right)  e^{-i\mathbf{q}\cdot\mathbf{r}}e^{i\omega t},\\
f\left(  \mathbf{r},t\right)   &  =\int\frac{d\mathbf{q}}{(2\pi)^{3}}\int%
\frac{d\omega}{2\pi}f\left(  \mathbf{q},\omega\right)  e^{i\mathbf{q}%
\cdot\mathbf{r}}e^{-i\omega t}. \label{EqInverseFT}%
\end{align}
Here the integration domain is a large system volume. In the free electron
gas
\begin{equation}
\chi\left(  q,\omega\right)  =\sum_{\mathbf{p}}\frac{\left(  f_{\mathbf{p}%
}-f_{\mathbf{p}+\mathbf{q}}\right)  \hbar^{2}/2}{\epsilon_{\mathbf{p}%
+\mathbf{q}}-\epsilon_{\mathbf{p}}+\hbar\omega+i0^{+}},
\end{equation}
where $f_{\mathbf{p}}=[\exp[(\epsilon_{\mathbf{p}}-\mu)/(k_{B}T)]+1]^{-1}$ is
the Fermi-Dirac distribution, $\epsilon_{\mathbf{p}}=\hbar^{2}p^{2}/\left(
2m_{e}\right)  $, $\mu$ is chemical potential, and $0^{+}$ is a positive
infinitesimal. The time constants of the conduction electrons in high density
metals are governed by the Fermi velocity (fs) and are much smaller than that
of the magnetization dynamics (ns), which justifies expansion to leading order
in the characteristic frequencies, i.e. the adiabatic
approximation~\citen{Simanek}, $\chi\left(  q,\omega\right)  \simeq\chi
_{r}(q)+i\omega\chi_{i}(q),$ where $\chi_{r}(q)=\lim_{\omega\rightarrow
0}\operatorname{Re}\chi\left(  q,\omega\right)  $ and $\chi_{i}(q)=\lim
_{\omega\rightarrow0}\partial_{\omega}\operatorname{Im}\chi(q,\omega).$ In the
three-dimensional free electron gas, the real part of the static
susceptibility $\chi_{r}(r)$ and its Fourier transform $\chi_{r}(q)$
correspond to the static RKKY and Lindhard functions
\begin{align}
\chi_{r}(r)  &  =\frac{g_{e}\hbar^{2}}{16\pi r^{3}}\left(  \frac{\sin2k_{F}%
r}{2k_{F}r}-\cos2k_{F}r\right)  ,\label{EqXrRKKY}\\
\chi_{r}(q)  &  =\frac{g_{e}\hbar^{2}}{8}\left(  1+\frac{k_{F}^{2}-(q/2)^{2}%
}{k_{F}q}\ln\left\vert \frac{k_{F}+q/2}{k_{F}-q/2}\right\vert \right)  ,
\label{EqXrRKKYQ}%
\end{align}
respectively~\citen{Lindhard}. The imaginary part of the susceptibility is
\begin{align}
\chi_{i}(r)  &  =\frac{g_{e}^{2}\hbar^{3}\pi}{8}\frac{\sin^{2}k_{F}r}%
{k_{F}^{2}r^{2}},\\
\chi_{i}(q)  &  =\frac{g_{e}^{2}\hbar^{3}\pi^{3}}{8k_{F}^{2}q}\Theta
(2k_{F}-q), \label{EqchiIq}%
\end{align}
where $k_{F}{\ =(3\pi^{2}n_{e})^{1/3}}$ is the Fermi wave number. Using
Eq.~(\ref{Sdrt})
\begin{equation}
\mathbf{s}_{c}(\mathbf{q},\omega)=\frac{J}{\hbar^{2}}\mathbf{S}(\omega
)\chi(q,\omega)\left[  n_{iso}(q)+n_{ani}(\mathbf{q})\right]  .
\label{Eqsc-momentum}%
\end{equation}
The Fourier transform of the density distribution $n_{d}=n_{iso}\left(
r\right)  +n_{ani}\left(  \mathbf{r}\right)  $ is the sum of
\begin{equation}
n_{iso}(q)=\left\langle j_{0}(qr)\right\rangle \label{EqnIso}%
\end{equation}
and%
\begin{align}
n_{ani}(\mathbf{q})  &  =-\pi\left\langle j_{2}(qr)\right\rangle Y_{z^{2}%
}\left(  \frac{\mathbf{q}}{q}\right)  \sqrt{\frac{5}{\pi}}\frac{Q_{2}}{\langle
r^{2}\rangle}\\
&  =-\frac{5Q_{2}}{4\left\langle r^{2}\right\rangle }\left\langle
j_{2}(qr)\right\rangle \left(  3\cos^{2}\theta_{\mathbf{q}}-1\right)  ,
\label{EqnAniCond}%
\end{align}
with $\cos\theta_{\mathbf{q}}=\mathbf{q}\cdot\mathbf{\hat{z}}$ and
$\left\langle j_{n}(qr)\right\rangle $ is the expectation value of the $n$-th
spherical Bessel function for a radial 3d wave function. Explicit formulas for
$\left\langle j_{0}(qr)\right\rangle $ and $\left\langle j_{2}%
(qr)\right\rangle $ are demoted to the appendix~\ref{AppendixSphericalBessel}.

\begin{figure}[tbh!]
\includegraphics[width=\columnwidth]{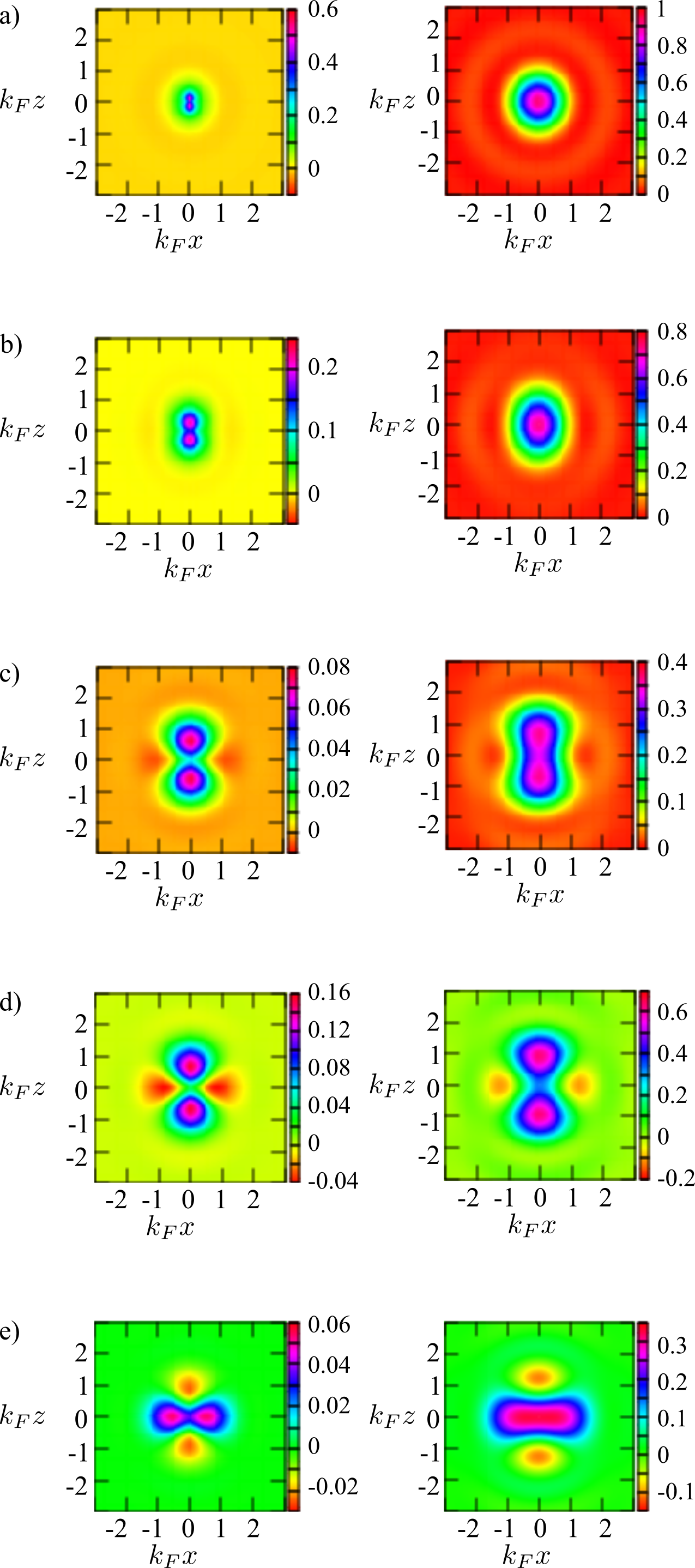}\caption{
Conduction electron spin densities induced by a time-dependent anisotropic
magnetic moment. $\rho_{r}(\mathbf{r})$ (left panels) and $\rho_{i}%
(\mathbf{r})$ (right panels) as defined in Eq.~(\ref{EqSpinDensityCE}) are
plotted for several parameter values in the $y=0$ plane. $\rho_{r}%
(\mathbf{r})$ and $\rho_{i}(\mathbf{r})$ are normalized to $2k_{F}^{2}g_{e}$
and $\chi_{i}(r=0)$, respectively. The values of $\left(  k_{F}^{2}%
\left\langle r^{2}\right\rangle ,Q_{2}/\left\langle r^{2}\right\rangle
\right)  $ are a)~$\left(  1/4,1/7\right)  $, b)~$\left(  1,1/7\right)  $,
c)~$\left(  4,1/7\right)  $, d)~$\left(  4,8/21\right)  $, e)~$\left(
4,-1/7\right)  $.}%
\label{Figspindensity}%
\end{figure}

Substituting $\chi\left(  q,\omega\right)  \simeq\chi_{r}(q)+i\omega\chi
_{i}(q)$ and keeping only linear terms in the frequency $\omega$ (adiabatic
approximation)
\begin{equation}
\mathbf{s}_{c}(\mathbf{q},\omega)=\frac{J}{\hbar^{2}}\left[  \mathbf{S}%
(\omega)\chi_{r}\left(  q\right)  +i\omega\mathbf{S}(\omega)\chi_{i}\left(
q\right)  \right]  n_{d}\left(  \mathbf{q}\right)  .
\end{equation}
Transforming back into time domain
\begin{equation}
\mathbf{s}_{c}(\mathbf{r},t)=\frac{J}{\hbar^{2}}\left[  \rho_{r}%
(\mathbf{r})\mathbf{S}(t)-\rho_{i}(\mathbf{r})\dot{\mathbf{S}}(t)\right]  .
\label{EqSpinDensityCE}%
\end{equation}
The densities
\begin{align}
\rho_{r}(\mathbf{r})  &  =\int\frac{d\mathbf{q}e^{i\mathbf{q}\cdot\mathbf{r}}%
}{(2\pi)^{3}}\chi_{r}(q)n_{d}(\mathbf{q})\\
\rho_{i}(\mathbf{r})  &  =\int\frac{d\mathbf{q}e^{i\mathbf{q}\cdot\mathbf{r}}%
}{(2\pi)^{3}}\chi_{i}(q)n_{d}(\mathbf{q})
\end{align}
are plotted in Fig.~\ref{Figspindensity} for several values of $k_{F}%
^{2}\langle r^{2}\rangle$ and $Q_{2}$. Figures~\ref{Figspindensity}a) and c)
illustrate that with increasing Fermi energy a larger region of the the
electron gas is polarized, as in the RKKY polarization
function~(\ref{EqXrRKKY}). The ion anisotropy is parameterized by the
quadrupole $Q_{2}$, which is proportional to $\langle r^{2}\rangle$, see
Eq.~(\ref{quadrupolev2}); larger ions induce a stronger anisotropy, cf.
Figs.~\ref{Figspindensity}c) and d). This can also be seen from
Eq.~(\ref{EqnAniCond}) by approximating $\langle j_{2}(qr)\rangle\approx
q^{2}\langle r^{2}\rangle/15,$ which leads to $n_{ani}\sim\langle r^{2}%
\rangle$. The sign of $Q_{2}$ can enlarge or decrease the total conduction
electron spin polarization, as shown in Figs.~\ref{Figspindensity}c) and e).

When the atomic radius is small $n_{d}(\mathbf{r})\rightarrow\delta
(\mathbf{r})$, the static spin polarization reduces to the well-known RKKY
spatial oscillations
\[
\lim_{n_{d}(\mathbf{r})\rightarrow\delta(\mathbf{r})}\rho_{r}(\mathbf{r}%
)=\chi_{r}(r),
\]
while $\rho_{i}(\mathbf{r})\rightarrow\chi_{i}(r)$. In this limit all crystal
field effects vanish. 

\section{Spin current}

\label{Sec:spincurrent} 
\begin{figure}[t!]
\includegraphics[width=\columnwidth]{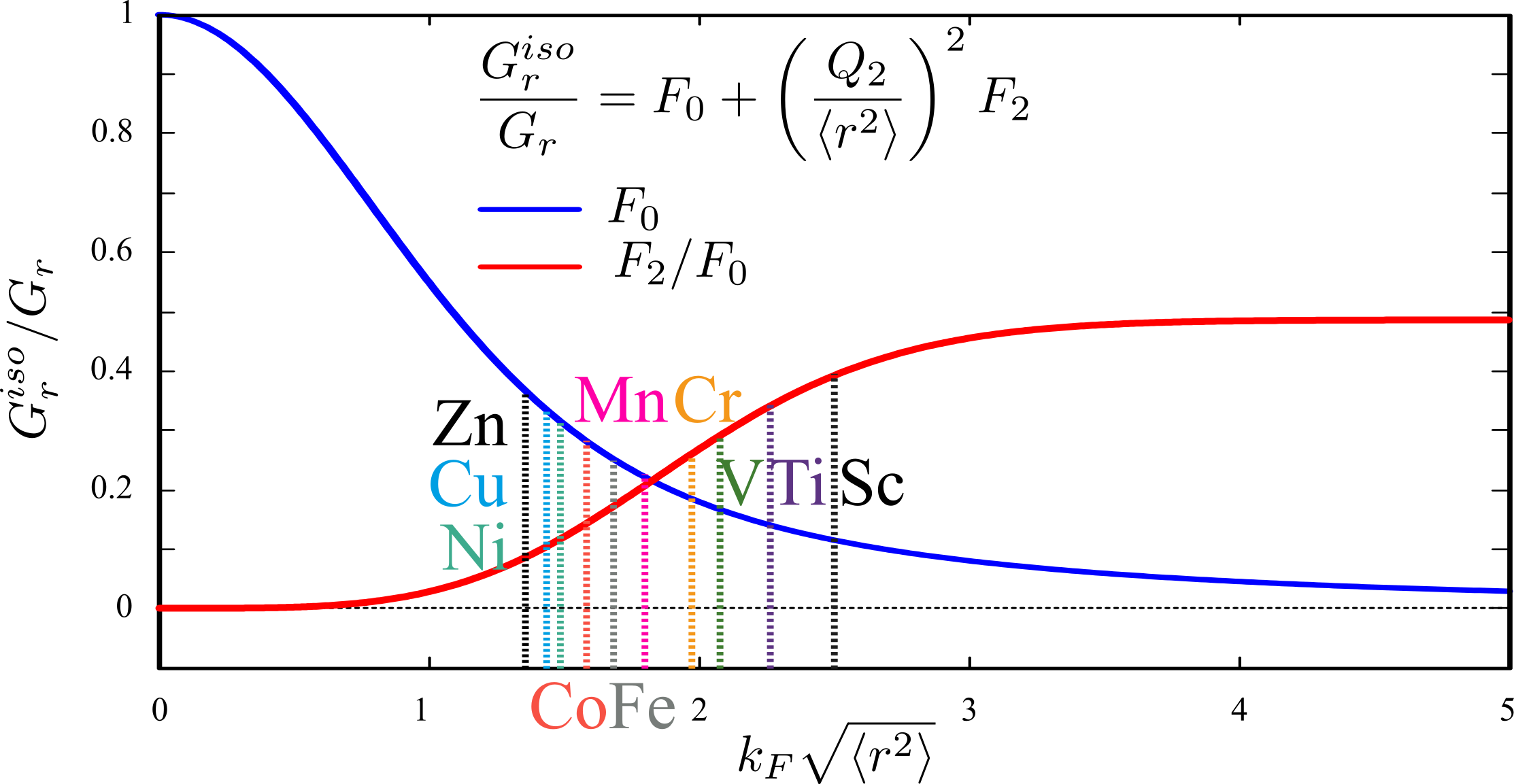}\caption{ Real
(dissipative) part of the spin mixing conductance $G_{r}^{iso}/G_{r}$ as
defined in Eqs.~(\ref{GRiso},\ref{Gr}) for 3d local moments in a free electron
metal with Fermi number $k_{F}\sim2$ \AA $^{-1}$ as a function of
$\left\langle r^{2}\right\rangle $, the mean square 3d orbital radius. The
suppression of $F_{0}$ with increasing $\left\langle r^{2}\right\rangle $
reflects the reduced Fourier components of exchange scattering at the Fermi
surface. Also indicated are the average 3d radii of free transition metal
atoms \protect\citen{waber} that decrease with higher nuclear charge. }%
\label{FigGR}%
\end{figure}

Conduction electron spin density and local moments are
also related by the spin conservation equation
\begin{equation}
\partial_{t}\mathbf{s}_{c}(\mathbf{r},t)+\nabla\cdot\mathbb{J}(\mathbf{r}%
)=\left(  \frac{d\mathbf{s}_{c}(\mathbf{r},t)}{dt}\right)  _{\mathrm{source}},
\end{equation}
where the source term
\begin{equation}
\left(  \frac{d\mathbf{s}_{c}(\mathbf{r},t)}{dt}\right)  _{\mathrm{source}%
}=-\frac{\mathbf{s}_{c}(\mathbf{r},t)}{\tau_{s}}+\frac{J}{\hbar^{2}}%
\mathbf{s}_{c}(\mathbf{r},t)\times\mathbf{S}_{d}(\mathbf{r},t),
\label{EqSourcespinconservation}%
\end{equation}
describes spin flip scattering on the time scale $\tau_{s}$ and spin
precession in the exchange torque exerted by the local moment. $\mathbb{J}%
_{\sigma}^{\nu}\mathbb{\ }$ is the spin current tensor, where the indexes
$\sigma$ and $\nu$ refer to the spin polarization and current directions,
respectively \citen{Chen}. We obtain explicit expressions for the spin current
divergence by substituting $\mathbf{s}_{c}$ from Eq.~(\ref{EqSpinDensityCE}),
in the clean limit of the metal and slow magnetization dynamics ($\tau
_{s}\rightarrow\infty$ and $\ddot{\mathbf{S}}\rightarrow0$):
\begin{equation}
\nabla\cdot\mathbb{J}(\mathbf{r})=\frac{J^{2}}{\hbar^{4}}\rho_{i}%
(\mathbf{r})n_{3d}\left(  \mathbf{r}\right)  \mathbf{S}\times\dot{\mathbf{S}%
}-\frac{J}{\hbar^{2}}\rho_{r}(\mathbf{r})\dot{\mathbf{S}}.
\label{spinconservation}%
\end{equation}
By writing the spin current in terms of a vector spin potential{\ }%
$\boldsymbol{\Phi}${$(\mathbf{r})$} as $\mathbb{J}_{\sigma}^{\nu}%
(\mathbf{r})=-\partial_{\nu}\boldsymbol{\Phi}(\mathbf{r})$,
Eq.~(\ref{spinconservation}) is reduced to a Poisson equation. The spin
current direction is governed by the gradient of the spin potential, while its
polarization is proportional to its direction. The solution of our Poisson
equation is%
\begin{equation}
\boldsymbol{\Phi}(\mathbf{r},t)=\Phi_{r}(\mathbf{r})\mathbf{S}(t)\times
\dot{\mathbf{S}}(t)+\Phi_{i}(\mathbf{r})\dot{\mathbf{S}}(t).
\end{equation}
where we defined dissipative ($\Phi_{r}$) and reactive ($\Phi_{i}$) scalar
potentials. To leading order in the quadrupole moment
\begin{equation}
\Phi_{r}(\mathbf{r})=\frac{G_{r}^{iso}}{4\pi r}\left(  1+\frac{3\cos^{2}%
\theta-1}{4r^{2}}Q_{2}\right)  ,
\end{equation}
where%
\begin{align}
{G_{r}^{iso}}  &  \equiv{\frac{J^{2}}{\hbar^{4}}\int\frac{d\mathbf{q}}%
{(2\pi)^{3}}\chi_{i}(q)\left\vert n_{d}(\mathbf{q})\right\vert ^{2}%
},\label{integralforGRiso}\\
&  =G_{r}\left[  F_{0}+\left(  \frac{Q_{2}}{\left\langle r^{2}\right\rangle
}\right)  ^{2}F_{2}\right]  , \label{GRiso}%
\end{align}
with
\begin{equation}
G_{r}=\frac{{\pi J^{2}g_{e}^{2}}}{{8\hbar}}. \label{Gr}%
\end{equation}
The dimensionless parameters can be obtained analytically as: {\footnotesize
\begin{align}
F_{0}  &  =\frac{11D(1208-5D(27D(3D-16)+682))-1627}{31185D(D+1)^{11}}%
+\frac{1627}{31185D},\\
F_{2}  &  =\frac{-44D(D(27D(12D-43)+985)+197)-788}{31185D(D+1)^{11}}%
+\frac{788}{31185D},
\end{align}
} where $D=k_{F}^{2}\left\langle r^{2}\right\rangle /14$, and are given in
Fig.~\ref{FigGR} for various transition metal atoms. With increasing ionic
radius, $F_{0}$ decreases, but $F_{2}$ increases up to half of $F_{0}$ for
lighter ions,{ because the ratio of the anisotropic contribution $\left\langle
j_{2}(qr)\right\rangle ^{2}/\left\langle j_{0}(qr)\right\rangle ^{2}$ is
suppressed for small $\left\langle r^{2}\right\rangle $ [see
Eq.~(\ref{EqBesselq})].}

When $k_{F}^{2}\left\langle r^{2}\right\rangle \rightarrow\infty$, both
$F_{0}$ and $F_{2}$ converge to zero as $\sim(k_{F}^{2}\left\langle
r^{2}\right\rangle )^{-1}$. While for small $k_{F}^{2}\left\langle
r^{2}\right\rangle \ll1$, $F_{0}\approx1$, $F_{2}\approx{4(k_{F}%
^{2}\left\langle r^{2}\right\rangle )^{2}}/{135}$ and $G_{r}^{iso}$ reduces to
$G_{r}$. $\Phi_{r}(\mathbf{r})$ decays monotonically with $r$, but with an
anisotropic component. The \textquotedblleft reactive\textquotedblright\ spin
potential in the \textquotedblleft far field\textquotedblright\ $r^{2}%
\gg\left\langle r^{2}\right\rangle $ reads to leading order in $r^{-1}$
\begin{equation}
\Phi_{i}(\mathbf{r})=\left(  \frac{1}{k_{F}^{2}}+Q_{2}\frac{3\cos^{2}\theta
-1}{3}\right)  \frac{G_{i}\cos2k_{F}r}{16\pi r^{3}},
\end{equation}
where $G_{i}={Jg_{e}}/{4}$. It oscillates as a function of distance as
$\cos(2k_{F}r),$ in phase with the RKKY-like ground state spin
density.

We can decompose the spin current along the radial $\hat{\mathbf{r}}$ and
polar $\hat{\boldsymbol{\theta}}$ unit vectors as
\begin{align}
\mathbf{J}^{r}(\mathbf{r})  &  =\frac{1}{4\pi r^{2}}\left(  G_{i}f_{2}%
(\theta)\frac{\sin2k_{F}r}{2k_{F}r}+f_{1}(\theta)G_{r}\mathbf{S}\times\right)
\dot{\mathbf{S}},\\
\mathbf{J}^{\theta}(\mathbf{r})  &  =\frac{3Q_{2}\sin\theta\cos\theta}{8\pi r^{4}%
}\left[  G_{i}\frac{\cos2k_{F}r}{3k_{F}r}-G_{r}\mathbf{S}\times\right]
\dot{\mathbf{S}}. \label{eqjI-ani}%
\end{align}
respectively, where $f_{1}(\theta)=1+3Q_{2}(3\cos^{2}\theta-1)/(4r^{2})$,
$f_{2}(\theta)=1+(2/3)k_{F}^{2}Q_{2}(3\cos^{2}\theta-1)$. The azimuthal
$\mathbf{J}^{\varphi}$ vanishes by symmetry. Ions with half-filled shells are
spherically symmetric and pump a radially symmetric spin current, i.e.
$\mathbf{J}^{\theta}\equiv0$.

The theory as exposed above is directly applicable to magnetic impurities in a
metal host. It induces anisotropy into the RKKY interaction between magnetic
moments in dilute alloys, which can be relevant for the Kondo and related
effects. Here we do not pursue this direction, since we are mainly interested
in the dynamics of interfaces between magnetic insulators and metals.

In transition metal oxides, magnetic cations usually fill the voids created by
oxygen anions scaffolding, with commonly tetrahedral and octahedral
coordination. In order to generate finite $Q_{2},$ the symmetry must be broken
by, e.g., by strain or at interface. This effect is at least partly
responsible for the large interface (compared to bulk) magnetic anisotropy of
transition metals \citen{Dieny}. 

\section{Local magnetic moment dynamics}

\label{Sec:damping} The spin current emitted by a local moment implies angular
momentum loss, that is, a dissipative torque acting on the local moment. In
the Landau-Lifshitz-Gilbert equation
\begin{equation}
\mathbf{\dot{M}}=-\gamma_{\mathrm{eff}}\mathbf{M}\times\mathbf{B}+\frac
{\alpha_{\mathrm{eff}}}{M_{s}}\mathbf{M}\times\mathbf{\dot{M}},
\end{equation}
spin pumping torques affect the gyromagnetic ratio $\gamma_{\mathrm{eff}}$ and
enhance the Gilbert damping $\alpha_{\text{\textrm{eff}}}$,
\begin{equation}
\gamma_{\mathrm{eff}}=\frac{\gamma_{0}}{1+G_{i}},\ \alpha_{\mathrm{eff}}%
=\frac{\alpha_{0}+M_{s}G_{r}^{iso}/\gamma_{0}}{\gamma_{0}/\gamma
_{\mathrm{eff}}}, \label{EqGyromagnetic}%
\end{equation}
where $\mathbf{M}$ is the magnetization vector, $\left\vert \mathbf{M}%
\right\vert =M_{s}$ is the saturation magnetization, and $\mathbf{B}$ is the
sum of external and anisotropy fields acting on the moment. The constants
$\gamma_{0}$ and $\alpha_{0}$ are the gyromagnetic ratio and Gilbert damping
in the absence of spin pumping, respectively. The anisotropic spin pumping
currents are not manifest in the magnetization dynamics because their torques
vanish when integrated over the local moment. $G_{r}^{iso}$ and $G_{i}$ play
roles equivalent to the real and imaginary part of the spin mixing conductance
at interfaces \citen{Tserkovnyak}. $G_{r}^{iso}$ parameterizes the dissipative
angular momentum and energy loss implied by spin pumping, just as the real
part of the spin mixing conductance at interfaces.

The imaginary part $G_{i}$ is sometimes referred to as an \textquotedblleft
effective magnetic field\textquotedblright. It apparently accelerates or
decelerates the precessional motion but \textit{conserves energy}. The present
results offer a simple picture of the physics of $G_{i}$ that has escaped
attention because it is hidden in the scattering theory formulation of spin
pumping: the coherent motion of the proximity RKKY spin density is locked to
the precessing magnetization of the local moment. {The Zeeman energy of the
uncoupled system acts only on the local magnetic moments
\begin{equation}
H_{Z}^{\left(  0\right)  }=\gamma_{0}\mathbf{B}\cdot\int\mathbf{S_{d}%
}(\mathbf{r},t)d\mathbf{r}=\gamma_{0}\mathbf{B}\cdot\mathbf{S}(t).
\end{equation}
On the other hand, it is the entire magnetic moment including the screeing
spins that precesses
\begin{equation}
\mathbf{M}=-\gamma_{0}\int\left[  \mathbf{S_{d}}(\mathbf{r},t)+\mathbf{s}%
_{c}(\mathbf{r},t)\right]  d\mathbf{r},
\end{equation}
where, in the adiabatic limit,
\begin{equation}
\int\mathbf{s}_{c}(\mathbf{r},t)d\mathbf{r}=\mathbf{S}(t)\frac{J}{\hbar^{2}%
}\int\rho_{r}(\mathbf{r})d\mathbf{r}=G_{i}\mathbf{S}(t),
\end{equation}
so $\mathbf{M}(t)=-\gamma_{0}\left(  1+G_{i}\right)  \mathbf{S}(t)$. The
Zeeman energy of the coupled systems therefore reads}%
\begin{equation}
H_{Z}=\frac{1}{1+G_{i}}\mathbf{B}\cdot\mathbf{M}(t).
\end{equation}
The renormalization field $\mathbf{B}\rightarrow\mathbf{B}\left(
1+G_{i}\right)  ^{-1}$ is therefore caused by the magnetic screening cloud
therefore that can equivalently be written in terms of a new gyromagnetic
ratio $\gamma_{0}\rightarrow\gamma_{\mathrm{eff}}\equiv\gamma_{0}\left(
1+G_{i}\right)  ^{-1}$.

\section{Magnetic insulator/normal metal interface}

\label{Sec:interface} The present results are relevant for an understanding of
the anisotropy at interfaces between normal metals and
ferromagnetic/ferrimagnetic insulators~\citen{AboutMetals}, such as garnets
and ferrites. The magnetism is then carried by local atomic moments that are
ordered by superexchange interactions, usually via oxygen anions. Since
localized on an atomic scale, only moments directly at the interface have a
significant exchange interaction with the conduction electrons in the metal.
Depending on the crystal direction and the interface cut, the number of
contributing magnetic moments varies, as does the spin mixing conductance
\citen{Jia}. Here we focus on the effects of the crystal field on the spin
pumping and the interface spin mixing conductance. The interface can be
modelled in terms of independent local moments~\citen{Jia} whose motion is
locked by the exchange coupling. The results for the single moments discussed
above can then be applied. Cubic sites, such as symmetric octahedrals, do not
deform the 3d electron density and suppress all anisotropies in cubic
ferromagnets. However, at the interface the bulk point symmetry is broken and
deformations normal to the interfaces may be expected, although we could not
find estimates for the magnitude of such interface crystal fields.

\begin{figure}[b!]
\includegraphics[width=\columnwidth]{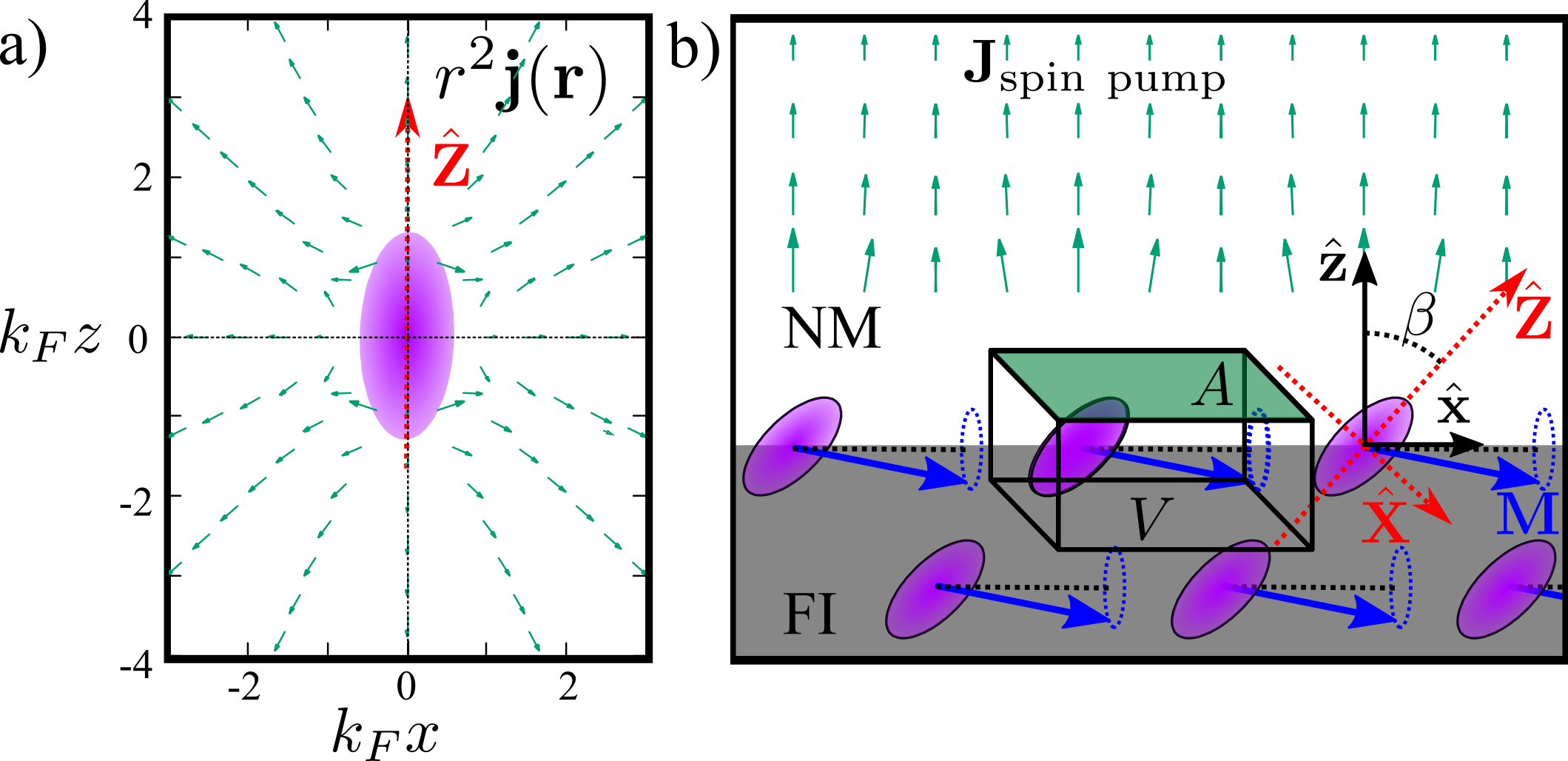}\caption{
Anisotropic spin pumping. a) Dissipative spin current pumped by a single
magnetic moment. Far from the origin, the spin current becomes isotropic. b)
Dissipative spin current generated in a bilayer of a ferromagnetic insulator
(FI) and normal metal (NM). Far from the interface the spin current direction
is normal to the interface. c) A sheet of magnetic moments is a model for the
FI$|$NM interface. }%
\label{ion}%
\end{figure}

Following Ref.~\citen{Simanek}, we model the metallic contact as a sheet of
magnetic ions in a free electron gas, see Fig.~\ref{ion}b). Spins can be
pumped only in one direction, so we are only interested in the results for
$z>0,$ i.e. the metallic side. We introduce the angle $\beta$ that
between the interface normal and the
local symmetry axes, see Fig.~\ref{ion}b). For example, when the crystal
surface is in (001) and (111) directions, the local symmetry axis is tilted by
angles $\beta=0$ and $\beta\simeq55^{\circ}$, respectively. The equilibrium magnetization is assumed to lie in the
interface by the thin-film easy-plane form anisotropy that is taken to
dominate any perpendicular crystalline magnetic anisotropy. The coordinate
along the interface normal $z$ is, in general, not parallel to the coordinate
$Z$ that points along the local crystal symmetry axis, see Fig.~\ref{ion}. We
adapt Eq. (\ref{Hsd}) to model the exchange interaction at an interface
\begin{equation}
H_{\mathrm{s-d}}=-\frac{J}{\hbar^{2}}\sum_{n}\int\mathbf{s}_{c}(\mathbf{r}%
,t)\cdot\mathbf{S}_{d}(\mathbf{r}-\mathbf{r}_{n},t)d\mathbf{r}
\label{hamilton-semi}%
\end{equation}
where the moments are at $\mathbf{r}_{n}=\left(  x_{n},y_{n},0\right)  $ in
the interface plane and $\mathbf{S}_{d}(\mathbf{r},t)$ has been defined in
Eq.~(\ref{Sdrt}). { Under FMR conditions all moments precess in phase. We
expand the magnetic moment density at the interface into plane waves with
reciprocal lattice vectors $\mathbf{G}=G_{x}\mathbf{\hat{\mathbf{x}}}%
+G_{y}\mathbf{\hat{y}}$
\begin{align}
\sum_{n}n_{d}\left(  \mathbf{r}-\mathbf{r}_{n}\right)   &  =\frac{N_{d}}%
{A}\sum_{\mathbf{G}}e^{i\mathbf{G}\cdot\left(  x\mathbf{\hat{\mathbf{x}}%
}+y\mathbf{\hat{y}}\right)  }\nonumber\\
&  \times\int\frac{dq_{z}}{2\pi}n_{d}(q_{z}\mathbf{\hat{z}}+\mathbf{G}%
)e^{iq_{z}z}, \label{EqDensitiesForTheLat3}%
\end{align}
where $N_{d}$ is the number of magnetic ions. The proximity conduction
electron spin density $\mathbf{s}_{c}(\mathbf{r},t)$ in linear response is
$\mathbf{s}_{c}(\mathbf{r})={J}{\hbar^{-2}}\left[  \rho_{r}(\mathbf{r}%
)\mathbf{S}(t)-\rho_{i}(\mathbf{r})\dot{\mathbf{S}}(t)\right]  $, where the
densities $\rho_{r,i}$ are also periodic in the interface plane,
\begin{align}
\rho_{r,i}(\mathbf{r})  &  =\frac{N_{d}}{A}\sum_{\mathbf{G}}e^{i\mathbf{G}%
\cdot\left(  x\mathbf{\hat{\mathbf{x}}}+y\mathbf{\hat{y}}\right)  }\int%
\frac{dq_{z}e^{iq_{z}z}}{2\pi}\nonumber\\
&  \times\chi_{r,i}(\sqrt{q_{z}^{2}+\mathbf{G}^{2}})n_{d}(q_{z}\mathbf{\hat
{z}}+\mathbf{G}). \label{EqDensitiesForTheLat2}%
\end{align}
}

The spin conservation equation in the metal reads
\begin{align}
\nabla\cdot\mathbb{J}  &  =\frac{J^{2}}{\hbar^{4}}\mathbf{S}(t)\times
\dot{\mathbf{S}}(t){\rho}_{i}(\mathbf{r})\sum_{n}n_{d}\left(  \mathbf{r}%
-\mathbf{r}_{n}\right) \nonumber\\
&  -\frac{J}{\hbar^{2}}\dot{\mathbf{S}}(t){\rho}_{r}(\mathbf{r}).
\label{spinconservationsemi}%
\end{align}
{In Appendic \ref{Sec:nonzeroG} we show that the in-plane components of the
spin current are exponentially suppressed with distance from the interface
with typical decay length of the order of the inverse of the (primitive)
reciprocal vector, $G^{-1}$, which can be estimated as $1\mathrm{\ nm}%
/(2\pi)=1.6\mathrm{\ \mathring{A}}$, for a lattice constant parameter of 1 nm.
The net spin current flow that leaves the magnet is therefore normal to the
interface direction as illustrated in Fig.~\ref{ion}b). }

The pumped spin current can be calculated by applying the Gauss theorem to a
flat \textquotedblleft pill box\textquotedblright\ with volume $V=Az$ as shown
in Fig.~\ref{ion}b). The spin current in the insulator vanishes, so
\begin{equation}
\int_{V}\nabla\cdot\mathbb{J}d\mathbf{r}=\mathbf{J}\int_{A}dxdy,
\label{EqGauss}%
\end{equation}
where $A$ is a surface on the metal side at a distance $z$ parallel to the
interface and $\mathbf{J/}\left\vert \mathbf{J}\right\vert $ is the current
polarization. {Then,
\begin{align}
\mathbf{J}\int_{B}dxdy  &  =\frac{J^{2}}{\hbar^{4}}\mathbf{S}(t)\times
\dot{\mathbf{S}}(t)\int_{V}d\mathbf{r}\sum_{n}{\rho}_{i}(\mathbf{r}%
)n_{d}\left(  \mathbf{r}-\mathbf{r}_{n}\right) \nonumber\\
&  -\frac{J}{\hbar^{2}}\dot{\mathbf{S}}(t)\int_{V}d\mathbf{r}{\rho}%
_{r}(\mathbf{r}). \label{EqGaussB}%
\end{align}
which has the solution
\begin{equation}
\mathbf{J}=g_{r}^{\uparrow\downarrow}\mathbf{m}\times\dot{\mathbf{m}}%
-g_{i}^{\uparrow\downarrow}{f}(z)\dot{\mathbf{m}},
\end{equation}
where
\begin{align}
g_{r}^{\uparrow\downarrow}  &  =\frac{N_{d}J^{2}S^{2}}{A\hbar^{4}}%
\sum_{\mathbf{G}}\int\frac{d\mathbf{q}}{(2\pi)^{3}}\chi_{i}(\sqrt
{q^{2}+\mathbf{G}^{2}})\left\vert n_{d}(q\mathbf{\hat{z}}+\mathbf{G}%
)\right\vert ^{2},\label{grup}\\
g_{i}^{\uparrow\downarrow}  &  =\frac{{N_{d}SG_{i}}}{{A}},
\end{align}
and the function $f(z)$ represents the integrated RKKY density in Fig.
~\ref{FigGI} that can be expressed analytically for short and long distances
$z$ from the interface:
\begin{align}
{f}\left(  z\gg k_{F}^{-1}\right)   &  \simeq\frac{\cos2k_{F}z}{\left(
2k_{F}z\right)  ^{2}}\left(  1-2Q_{2}k_{F}^{2}\frac{3\cos^{2}\beta-1}%
{3}\right)  ,\label{fz}\\
{f}\left(  z\ll k_{F}^{-1}\right)   &  \simeq1+2Q_{2}k_{F}^{2}\frac{3\cos
^{2}\beta-1}{9}. \label{fz2}%
\end{align}
}
 \begin{figure}[t!]
\includegraphics[width=\columnwidth]{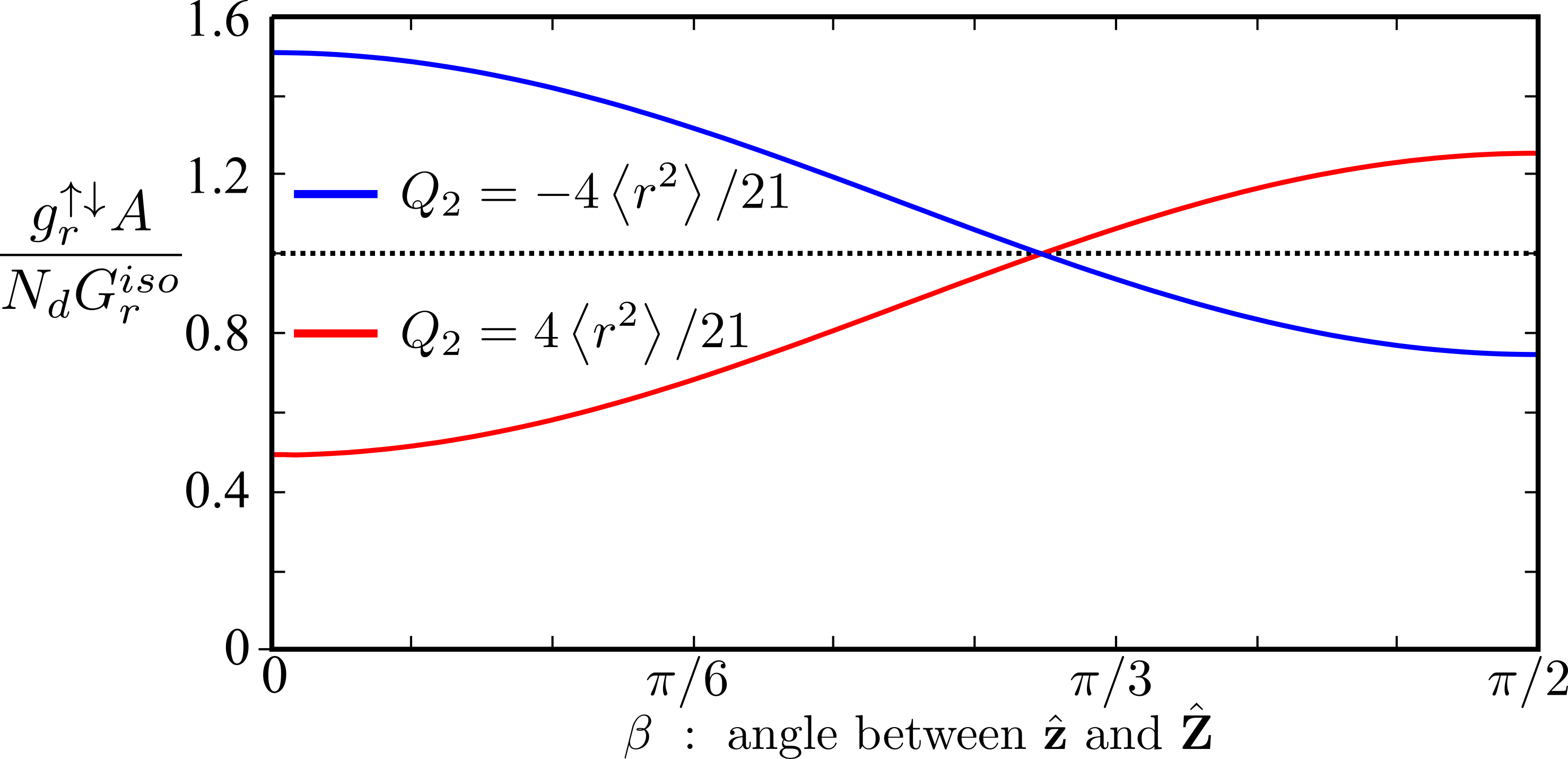}\caption{ The
dissipative spin current injected by a ferromagnetic insulator into a normal
metal $g_{r}^{\uparrow\downarrow}$ as a function of angle $\beta$ between
crystal field direction and interface normal. }%
\label{FigGRtheta}%
\end{figure}

{The coefficients $g_{r}^{\uparrow\downarrow}$ and $g_{i}^{\uparrow\downarrow
}$ as obtained by integrating the right-hand-side of
Eq.~(\ref{spinconservationsemi}) represent the real and imaginary parts of the
spin-mixing conductance, respectively. The sum over }$\mathbf{G}${ in Eq.
(\ref{grup}) reflects interference effects }that can be simplified in the
limit of large density of magnetic moments, i.e. when $a\leq\pi/k_{F}$. Since
the susceptibility $\chi_{i}(q)$ is proportional to the step function, see
Eq.~(\ref{EqchiIq}), and modes with wavenumber $\left\vert \mathbf{G}%
\right\vert \geq2k_{F}$ {do not contribute. In that limit
\begin{align}
g_{r}^{\uparrow\downarrow}  &  \approx\frac{N_{d}J^{2}S^{2}}{A\hbar^{4}}%
\int\frac{d\mathbf{q}}{(2\pi)^{3}}\chi_{i}(q)n_{d}(q\hat{\mathbf{z}}%
)n_{d}(-q\hat{\mathbf{z}})\label{EqGrFinal2}\\
&  =\frac{N_{d}S^{2}G_{r}^{iso}}{A}\left(  1-Q_{2}k_{F}^{2}\frac{3\cos
^{2}\beta-1}{3}\right)  , \label{EqaniJ}%
\end{align}
where $G_{r}^{iso}$ is given by Eq. (\ref{GRiso}). The contribution by Bragg
scattering with finite $G$ is relevant for materials with lower interface
moment density. However, Eq. (\ref{grup}) as a function of the interface
moment density can be calculated only numerically, but we estimate that
correction terms suppress the anisotropy. In the dilute limit }$N_{d}%
\rightarrow0${ all interference and thereby anisotropies vanish, as discussed
in Appendix~\ref{AppendixFiniteG}. }

For a $3d^{6}$ high-spin state of Fe$^{2+}$ or Co$^{3+}$: $S=3\hbar/2$ and
$N_{d}/A=(0.5\,\mathrm{nm})^{-2}$, the isotropic contribution {${g}%
_{r}^{\uparrow\downarrow}=N_{d}S^{2}G_{r}^{iso}/A\sim10^{18}\,$ cm$^{-2}$}%
$\ $is of the order of magnitude accepted for magnetic insulators, while{
$g_{i}^{\uparrow\downarrow}=N_{d}SG_{i}/A\sim10^{18}$ cm$^{-2}$ appears to be
rather large }\citen{Weiler}. The anisotropy of the spin current pumped by
transition metal ions in an elongated octahedral crystal field and high spin
state with $Q_{2}=\pm4/21$ is plotted in Fig.~\ref{FigGRtheta}. The magnetic
ions emits less spin current in the direction in which the 3d sub shell is
elongated,{ because the spin current is generated by exchange in the overlap
volume of the spin densities $\mathbf{s_{c}}$ and $\mathbf{S_{d}}$, see
Eqs.~(\ref{spinconservation}) and~(\ref{EqSourcespinconservation})}, and
{$\mathbf{s_{c}}$ is preferentially suppressed{ by proximity} in that
direction}. Ions with positive $Q_{2}$ generate maximal (minimal) spin current
for $\beta=\pi/2$ $\left(  \beta=0\right)  $, while this is opposite for
negative $Q_{2}$. The anisotropic spin current depend on the relative angle
between the $3d$ sub shell orientation and the interface orientation $\beta,$
which should be observable at selected interfaces.

The reactive spin current depends on position by the function $f(z).$ It is
not a tranport spin current, but cause by the coherent precession of the
proximity spin density. It can be obtained by applying Gauss' theorem to an
integral over the volume $V=Az$, as in Fig.~\ref{FigGI}. The magnitude of this
current near the interface is estimated $g_{i}^{\uparrow\downarrow}%
\simeq10^{18}$ cm$^{-2}$ and vanishes algebraically with distance from the
interface with the RKKY oscillations. This spatial dependence $f(z)$ is lost
in scattering theory in which only the transport of electrons between
incoherent reservoirs are considered. The imaginary part of the spin mixing
conductance $g_{i}^{\uparrow\downarrow}$ has been found to be relatively small
for most systems~\citen{SpinMixinConductance2005}.
\begin{figure}[t!]
\includegraphics[width=\columnwidth]{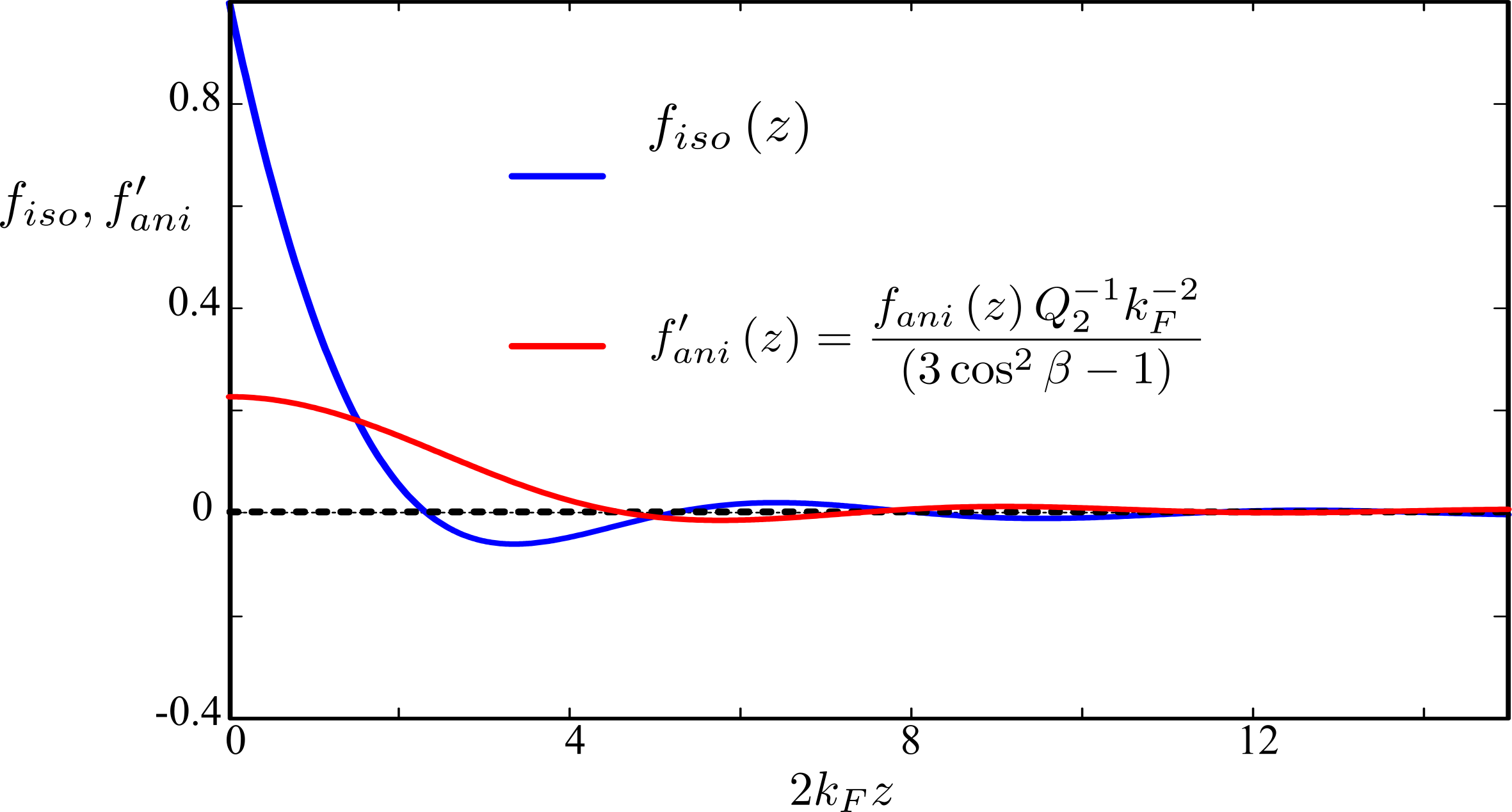}\caption{ The
RKKY-like spatial oscillation, represented by dimensionless $f(z)$ in the
reactive spin current $g_{i}^{\uparrow\downarrow}{f}(z)$, where ${f}={f}%
_{iso}+{f}_{ani}$ is a decomposed into an isotropic (${f}_{iso}$) and
anisotropic (${f}_{ani}$) terms. $f_{ani}^{\prime}(z)=f_{ani}(z)Q_{2}%
^{-1}k_{F}^{-2}(3\cos^{2}\beta-1)^{-1}$ is the normalized $f_{ani}$. For $z\gg
k_{F}^{-1}$, ${f}_{iso}$ and ${f}_{ani}^{\prime}$ approaches $\left(
2k_{F}\right)  ^{-2}\cos2k_{F}z$ and $-(2/3)\left(  2k_{F}\right)  ^{-2}%
\cos2k_{F}z$, respectively. When $Q_{2}\neq0$ the spin current depends on the
angle $\beta$. }%
\label{FigGI}%
\end{figure}

Cobalt ferrite (CoFe$_{2}$O$_{4}$ or CFO) is an iron-based spinel. Cobalt
ferrites possess an inverse spinel structure, [Fe$^{3+}$]$_{\text{T}}%
$[Co$^{2+}$Fe$^{3+}$]$_{\text{O}}$O$_{4}$, where the subscripts []$_{T}$ and
[]$_{O}$ stand for the tetrahedral and octahedral sites, respectively. The
iron ions have half filled subshell and an isotropic electronic cloud,
regardless of the symmetries of their environment. However, the octahedrally
coordinated Co$^{2+}$ ions occupy elongated octahedrals when grown on
SrTiO$_{3}$ (STO) substrates. The unit cell lattice parameter of STO
$a_{\text{STO}}=3.906$ \AA ~\citen{Auxetic} is smaller than the corresponding
lattice parameter of CFO $a_{\text{CFO}}=4.195$ \AA ~\citen{Auxetic}. As a
result of this lattice mismatch, CFO films are in-plane compressed and
tetragonally distorted~\citen{Casanova,Auxetic2}, depending on the grown
direction of the sample. When CFO is grown in the (001) direction, the
resultant crystal field is an elongated octahedral, while in the (111) growth
direction the compression creates a slanted octahedral crystal field. The
resultant crystal field can be described by that of an elongated octahedral
with a small energy splitting (see Appendix~\ref{AppendixDistort}). Our model
predicts that the exchange between the cobalt ions and the conduction
electrons is stronger for a (001) CFO than for a (111) one. Indeed, replacing
the cobalt quadrupole $Q_{2}=-4/21$ in Eq.~(\ref{EqaniJ}) for the angles
$\beta_{(001)}=0$ and $\beta_{(111)}=\pi/4$, we find that $g_{r,(001)}%
^{\uparrow\downarrow}$ is 50 $\%$ larger than $g_{r,(111)}^{\uparrow
\downarrow}$, in agreement with the experiment of Ref.~\citen{Casanova}. It
should be mentioned that the magnetization and the surface Co$^{2+}$/Fe$^{3+}$
concentration ratio strongly depend on the preparation
conditions~\citen{Casanova}, however.

{Our model can be applied to other than ferromagnetic order of the local
moments at the interface. The dissipative spin current emitted by each ion is
proportional to $\mathbf{S}\times\mathbf{\dot{S}}$ and thereby invariant to
spin reversal $\mathbf{S}\rightarrow-\mathbf{S}$. The sum of all spin-current
contributions, as well as the real part of the spin mixing conductance, does
not depend on the (collinear) order of the sub-lattices (ferro, ferri or
antiferromagnetic) \citen{Jia}. However, the imaginary part of the spin mixing
conductance or effective exchange field $g_{i}^{\uparrow\downarrow}$ felt by
the conduction electrons is governed by the sum of the local moments and
vanishes for exactly compensated antiferromagnetic interface order. This is
consistent with previous studies of the spin-pumping by
antiferromagnets~\citen{SPAFM,Takei}. }

\section{Concluding remarks}

\label{SecConclusions}

The deformations of partially filled 3d shells of local moments in noncubic
crystal fields are reflected by anisotropic RKKY spin-density oscillations and
nonlocal exchange interactions in metallic hosts. We show that the spin
current pumped by a magnetic moment with nonspherical spin density is
anisotropic as well. The spin pumping leads to enhanced magnetization damping
and a renormalized gyromagnetic ratio. The latter can be interpreted in terms
of the coherent motion of the RKKY spin density oscillations.

The properties of interfaces between magnetic insulators and metals are
governed by the local moments in the terminating monolayer. The spin mixing
conductance and its asymmetry depends not only on the density of exposed
moments, but also on the local point symmetry. We applied the theory to
analyze the spin pumping from a ferromagnetic insulator to an adjacent normal
metal. Most anisotropies focus the spin currents into a direction normal to
the interface, which is beneficial for spintronics. { Spin pumping
and spin transfer torque are each others Onsager reciprocals and governed by
the same spin-mixing conductance. The crystal field effects in spin pumping
addressed here therefore equally affect the spin-transfer torque efficiency. 

The anisotropy of the pumped spin-current depends on the quadrupole moment
$Q_{2}$, which in turns depends on the orbital occupation of interface
magnetic atoms. While we focus here on CFO,  the anisotropy should affect
all transition metal based magnetic insulator with magnetic moments at the
interface with nonspherical spin distribution. An interesting material to
apply the present analysis could be La$_{1-x}$Sr$_{x}$MnO$_{3}$ (LSMO) in
which the 3d shell of the Mn ions is not half filled.} First principles band
structure calculations can test our predictions and render them more quantitative.

\textit{Acknowledgments.-} A.B.C. acknowledges a JSPS Fellowship for Young
Scientists {  No. JP15J02585}. This work was supported by JSPS
KAKENHI Grant Nos. 25247056, 25220910, 26103006 {  and CONICYT
Becas Chile 74170017}.

\appendix

\section{Tesseral spherical harmonics}

\label{AppendixYj} 
\begin{figure}[tbh!]
\begin{center}
\includegraphics[width=\columnwidth]{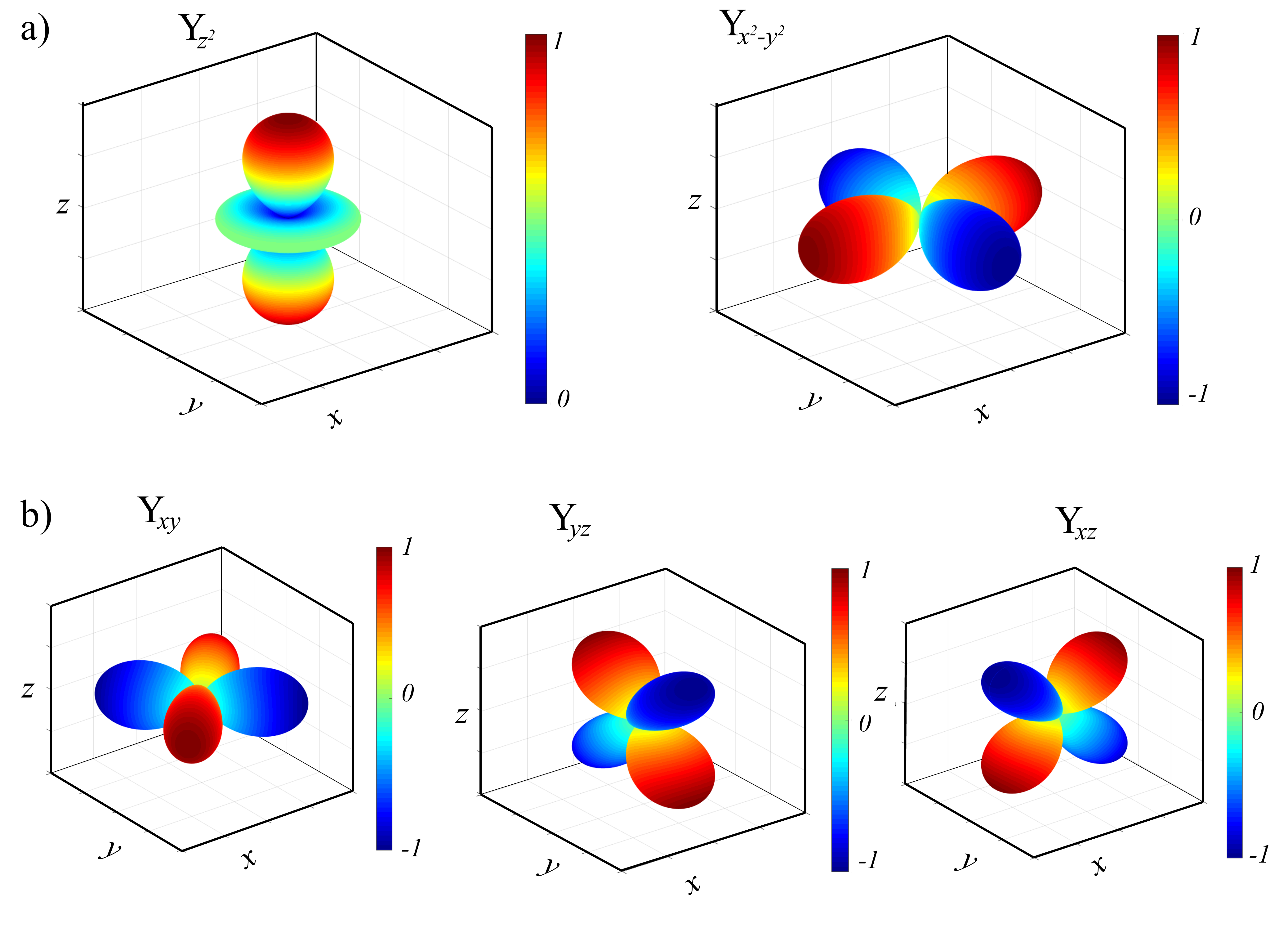}
\end{center}
\caption{	 Orbitals of the 3d atomic shell. For each plot, the
radius of the surface is the value of the function $Y_{j}$, $r(\theta
,\phi)=Y_{j}$, where $j$ is the orbital label. a) $e_{g}$ orbitals have lobes
along the crystal axes. b) $t_{2g}$ orbitals point between the axes, .}%
\label{Fig3dOrbitals}%
\end{figure}

The $3d$ orbitals can be written as
\begin{equation}
\psi_{j}(\mathbf{r})=R_{3d}(r)Y_{j}(\mathbf{\Omega}),
\end{equation}
where the radial function can be approximated by Slater-type wave
functions~\citen{freeman,waber},
\begin{equation}
R_{3d}(r)=\sqrt{\frac{1}{6!}\left(  \frac{2}{a}\right)  ^{7}}r^{2}e^{-r/a},
\end{equation}
where the constant $a$ is related to the mean-square radius by $\left\langle
r^{2}\right\rangle =14a^{2}\sim1$ \AA . In crystals, the angular part of the
$3d$ wavefunctions is described by the set of
orbitals~\citen{BookMagnetismBlundell}
\begin{align}
Y_{z^{2}}  &  =\sqrt{\frac{5}{16\pi}}\left(  3\cos^{2}\theta-1\right)
\nonumber\\
&  =\sqrt{\frac{5}{16\pi}}\frac{2z^{2}-x^{2}-y^{2}}{r^{2}},\\
Y_{x^{2}-y^{2}}  &  =\sqrt{\frac{15}{16\pi}}\sin^{2}\theta\cos2\phi\nonumber\\
&  =\sqrt{\frac{15}{16\pi}}\frac{x^{2}-y^{2}}{r^{2}},
\end{align}
known as $e_{g}$ orbitals, and
\begin{align}
Y_{xy}  &  =\sqrt{\frac{15}{16\pi}}\sin^{2}\theta\sin2\phi=\sqrt{\frac
{15}{4\pi}}\frac{xy}{r^{2}},\\
Y_{yz}  &  =\sqrt{\frac{15}{16\pi}}\sin2\theta\sin\phi=\sqrt{\frac{15}{4\pi}%
}\frac{yz}{r^{2}},\\
Y_{zx}  &  =\sqrt{\frac{15}{16\pi}}\sin2\theta\cos\phi=\sqrt{\frac{15}{4\pi}%
}\frac{zx}{r^{2}},
\end{align}
known as $t_{2g}$ orbitals. They are shown in Fig.~\ref{Fig3dOrbitals}.

\subsection{Mean value of spherical Bessel functions}

\label{AppendixSphericalBessel} The $n$-th spherical Bessel function for $n=0$
and $n=2$ read
\begin{align}
j_{0}(x)  &  =\frac{\sin x}{x},\\
j_{2}(x)  &  =\left(  \frac{3}{x^{2}}-\frac{1}{x}\right)  \frac{\sin x}%
{x}-\frac{3\cos x}{x^{2}}. \label{EqBesselDef}%
\end{align}
Their mean values over Slater-type single-exponential orbitals are
\begin{align}
\left\langle j_{0}(qr)\right\rangle  &  =\int j_{0}(qr)n_{d}(\mathbf{r}%
)d\mathbf{r}=\frac{1-\frac{5}{6}a^{2}q^{2}+\frac{1}{16}a^{4}q^{4}}{\left(
1+\frac{1}{4}a^{2}q^{2}\right)  ^{6}}\nonumber\\
&  \approx\left(  1-\frac{q^{2}\left\langle r^{2}\right\rangle }{6}\right) \\
\left\langle j_{2}(qr)\right\rangle =  &  \int j_{2}(qr)n_{d}(\mathbf{r}%
)d\mathbf{r}=\frac{2a^{2}q^{2}\left(  7-3a^{2}q^{2}/4\right)  }{15\left(
1+a^{2}q^{2}/4\right)  ^{6}}\nonumber\\
&  \approx\frac{q^{2}\left\langle r^{2}\right\rangle }{15}, \label{EqBesselq}%
\end{align}
where the approximations are valid for $q^{2}\left\langle r^{2}\right\rangle
\ll1$.

\section{Spin current direction for interfaces}

\label{Sec:nonzeroG} A two-dimensional periodic lattice is specified by two
independent primitive translation vectors, $\mathbf{a}_{1}$ and $\mathbf{a}%
_{2}$
\begin{equation}
\mathbf{r}_{\mathbf{n}}=n_{1}\mathbf{a}_{1}+n_{2}\mathbf{a}_{2}.
\end{equation}
Its reciprocal lattice is
\begin{equation}
\mathbf{G}=n_{1}^{\prime}\mathbf{b}_{1}+n_{2}^{\prime}\mathbf{b}_{2},
\end{equation}
where $\mathbf{b}_{1}=2\pi\mathbf{a}_{2}/\left\vert \mathbf{a}_{1}%
\times\mathbf{a}_{2}\right\vert $ and $\mathbf{b}_{2}=2\pi\mathbf{a}%
_{1}/\left\vert \mathbf{a}_{1}\times\mathbf{a}_{2}\right\vert $.
Equation.~(\ref{spinconservationsemi}) implies that translational symmetry of the
density $n_{d}$ and $\rho_{r}$ carries over to the vector spin potential
$\mathbf{\Phi}$ and spin current tensor $\mathbb{J}_{\sigma}^{\nu}%
(\mathbf{r})=-\partial_{\nu}\boldsymbol{\Phi}(\mathbf{r})$. Therefore
\begin{equation}
\boldsymbol{\Phi}(\mathbf{r})=\sum_{\mathbf{G}}\boldsymbol{\Phi}_{\mathbf{G}%
}(z)e^{i\left(  G_{x}x+G_{y}y\right)  }.
\end{equation}
where the Fourier coefficients $\boldsymbol{\Phi}_{\mathbf{G}}$ depends on the
distance $z$ from the interface. Inserting this expansion into
Eq.~(\ref{spinconservationsemi}) leads to
\begin{equation}
\left[  G^{2}-\partial_{z}^{2}\right]  \boldsymbol{\Phi}_{\mathbf{G}}%
(z)=\frac{1}{A}\int_{A}dxdye^{-i\left(  G_{x}x+G_{y}y\right)  }%
\boldsymbol{\mathcal{T}}(\mathbf{r}),
\end{equation}
with
\begin{equation}
\boldsymbol{\mathcal{T}}(\mathbf{r})=-\frac{J}{\hbar}{\rho}_{r}(\mathbf{r}%
)\dot{\mathbf{S}}+\frac{J^{2}}{\hbar^{4}}{\rho}_{i}(\mathbf{r})\sum_{n}%
n_{d}\left(  \mathbf{r}-\mathbf{r}_{n}\right)  \mathbf{S}\times\dot
{\mathbf{S}}.
\end{equation}
where $A$ is the unit cell area. Fourier transforming with respect to $z$
gives
\begin{align}
\left[  G^{2}+q_{z}^{2}\right]   &  \boldsymbol{\Phi}_{\mathbf{G}}%
(q_{z})=\boldsymbol{\mathcal{T}}_{\mathbf{G}}(q_{z}),\\
\boldsymbol{\mathcal{T}}_{\mathbf{G}}(q_{z})  &  =A^{-1}\int dze^{-iq_{z}%
z}\int dxdye^{-i\left(  G_{x}x+G_{y}y\right)  }\boldsymbol{\mathcal{T}%
}(\mathbf{r}).
\end{align}
Hence
\begin{align}
\boldsymbol{\Phi}(\mathbf{r})  &  =\sum_{\mathbf{G}}e^{i\left(  G_{x}%
x+G_{y}y\right)  }\int\frac{dq_{z}e^{iq_{z}z}}{2\pi(q_{z}^{2}+G^{2}%
)}\boldsymbol{\mathcal{T}}_{\mathbf{G}}(q_{z})\\
&  =\sum_{\mathbf{G}}e^{i\left(  G_{x}x+G_{y}y\right)  }\frac{e^{-Gz}}%
{2G}\boldsymbol{\mathcal{T}}_{\mathbf{G}}(iG),
\end{align}
using the residuals theorem for the pole $q_{z}=iG$. $\mathbb{J}$ can be
decomposed into the currents along $\mathbf{z}$ and $\mathbf{G}$ as{%
\begin{align}
\mathbb{J}^{z}  &  =\frac{1}{2}\sum_{\mathbf{G}}e^{-Gz}e^{i\left(
G_{x}x+G_{y}y\right)  }\mathbf{\hat{z}}\otimes\boldsymbol{\mathcal{T}%
}_{\mathbf{G}}(iG)\\
\mathbb{J}^{\mathbf{G}}  &  =\frac{-i}{2}\sum_{\mathbf{G}}e^{-Gz}e^{i\left(
G_{x}x+G_{y}y\right)  }\frac{\mathbf{G}}{G}\otimes\boldsymbol{\mathcal{T}%
}_{\mathbf{G}}(iG),
\end{align}
where $\otimes$ is the external product of the two subspaces (spin direction
and current flow direction). Thus the spin current flowing in the in-plane
directions ($\mathbf{J}^{\mathbf{G}}$) decays exponentially with distance $z$
when $\mathbf{G}\neq\mathbf{0}$. Only the contribution perpendicular to the
interface ($G=0$) propagates as
\begin{equation}
\mathbf{J}^{z}(z\gg\langle r\rangle)\sim A^{-1}\int d\mathbf{r}%
\boldsymbol{\mathcal{T}}(\mathbf{r}),
\end{equation}
which is the same result we obtain in the main text using the divergence
theorem. On the other hand, $\mathbf{J}^{\mathbf{G}}$ is not defined for
$G=0$. }\twocolumngrid

\section{Finite wavelength contributions to the uniform spin current}

\label{AppendixFiniteG} { Corrections for finite $G=\left\vert \mathbf{G}%
\right\vert <2k_{F}$ can be calculated by using the susceptibility $\chi
_{r}(q)$ and spin-density $n_{d}(\mathbf{q})$ in Eqs.~(\ref{EqchiIq}%
),~(\ref{EqnIso}) and~(\ref{EqnAniCond}) in the spin-mixing conductance
formula~(\ref{grup}). Equation~(\ref{grup}) can be rewritten in term of
lattice vector $\mathbf{a}_{n_{x},n_{y}}=n_{x}a_{x}+n_{y}a_{y}$,
\begin{equation}
g_{r}^{\uparrow\downarrow}=\frac{N_{d}J^{2}S^{2}}{A\hbar^{4}}\int%
\frac{d\mathbf{q}}{(2\pi)^{3}}\chi_{i}(q)\left\vert n_{d}(\mathbf{q}%
)\right\vert ^{2}\sum_{n_{x},n_{y}}e^{i\mathbf{q}\cdot\mathbf{a}_{n_{x},n_{y}%
}}. \label{EqG_rGeneralized}%
\end{equation}
The numerical integration of the above equation is tedious. However, two
natural limit cases are analytically accesible, namely the dense ($a_{x}%
,a_{y}\rightarrow0$) and dilute ($a_{x},a_{y}\rightarrow\infty$) local moments
approximations. While the former is used in the main text, we address here the
second one by the following expansion,
\[
e^{i\mathbf{q}\cdot\mathbf{a}_{n_{x},n_{y}}}=4\pi\sum_{lm}i^{l}j_{l}\left(
qa_{n_{x},n_{y}}\right)  Y_{lm}^{\ast}\left(  \hat{\mathbf{q}}\right)
Y_{lm}\left(  \hat{\mathbf{a}}_{n_{x},n_{y}}\right)  .
\]
For garnets with large lattice constants (large $a_{n}$), we can use the
asymptotic properties of the spherical Bessel functions,%
\begin{equation}
\lim_{qa_{n_{x},n_{y}}\gg1}j_{l}\left(  qa_{n_{x},n_{y}}\right)  =\frac
{\sin\left(  qa_{n_{x},n_{y}}-l\frac{\pi}{2}\right)  }{qa_{n_{x},n_{y}}}.
\end{equation}
Therefore the contribution of the $n_{x},n_{y}\neq0$ terms decays as
$\propto(k_{F}a_{n_{x},n_{y}})^{-1}$ and for a moment-to-moment distance of a
nm and an elemtental metal typically smaller than 0.1. The sum of
Eq.~(\ref{EqG_rGeneralized}) is then dominated by the isotropic $n_{x}%
=n_{y}=0$ term and the interface spin-mixing conductance is just the sum of
the (isotropic) single ion contributions, $g_{r}^{\uparrow\downarrow}%
=NG_{r}^{iso}$. The spin current generated by well-separated magnetic moments
does cause interference effects on the perpendicular spin current and the
anisotropies vanish.}

\section{Crystal field of distorted octahedral sites}

\label{AppendixDistort} { Growing CFO films on lattice mismatched
substrates~\citen{Casanova} causes magnetostriction~\citen{StrainCFO} that
leads to a distortion of the octahedral environment of the cobalt moments. An
elongation or contraction in the crystal directions (001) and (111) shifts the
oxygen ion positions along the direction $\left(  \frac{\sin\beta}{\sqrt{2}%
},\frac{\sin\beta}{\sqrt{2}},\cos\beta\right)  $, where $\beta_{(001)}=0$ and
$\beta_{(111)}=55^{\circ}$. Similar to Eq.~(\ref{Eq_cfpoints}), the point
charge model leads to a crystal field splitting }%
\begin{equation}
\Delta H_{cf}^{\beta}(\mathbf{r})=\Delta_{0}\frac{2z^{2}-x^{2}-y^{2}}{\langle
r^{2}\rangle},
\end{equation}
where
\begin{equation}
\Delta_{0}=\frac{3\left\langle r^{2}\right\rangle \delta\cos\left(
\beta+\beta_{0}\right)  }{4\pi\epsilon_{0}R_{0}^{3}\cos\beta_{0}},
\end{equation}
$\delta$ is the strain, and $\beta_{0}\simeq35^{\circ}$. This distortion
creates an effective quadrupolar crystal-field. The lattice constants of CFO,
SrTiO$_{3}$ substrate, and Pt overlayer are $a_{\mathrm{STO}}=3.906$ \AA ,
$a_{\mathrm{CFO}}=4.195$ \AA , $a_{\mathrm{Pt}}=3.912$ \AA \ \citen{Dhakal}.
With strain $\delta\sim2\%$, we estimate $\Delta_{0}\sim0.06$ eV.


\begin{thebibliography}{99}                                                                                               %


\bibitem {Tserkovnyak}Y. Tserkovnyak, A. Brataas and G. E. W. Bauer.
\textit{Enhanced Gilbert damping in thin ferromagnetic films}, Phys. Rev.
Lett. \textbf{88}, 117601 (2002).

\bibitem {Simanek}E. Simanek and B. Heinrich. \textit{Gilbert damping in
magnetic multilayers}, Phys. Rev. B \textbf{67}, 144418 (2003).

\bibitem {Brataas2012}A. Brataas, Y. Tserkovnyak, G.E.W. Bauer, and P.J.
Kelly, in \textit{Spin Current}, edited by S. Maekawa, E. Saitoh, S.
Valenzuela and Y. Kimura (Oxford University Press, New York, 2012) p. 87-135,

\bibitem {Jia}X. Jia, K. Liu, K. Xia and G.E.W. Bauer, \textit{Spin transfer
torque on magnetic insulators}, Europhys. Lett. \textbf{96}, 17005 (2011).

\bibitem {Burrowes}C.~Burrowes, B.~Heinrich, B.~Kardasz, E.A. Montoya,
E.~Girt, Y.~Sun, Y.-Y.~Song, and M.~Wu. \textit{Enhanced spin pumping at
yttrium iron garnet/Au interfaces}, Appl. Phys. Lett. \textbf{100}, 092403 (2012).

\bibitem {Weiler}M.~Weiler, M.~Althammer, M.~Schreier, J.~Lotze,
M.~Pernpeintner, S.~Meyer, H.~Huebl, R.~Gross, A.~Kamra, J.~Xiao, Y.~Chen,
H.~Jiao, G.~E.~W.~Bauer, and S.~T.~B.~Goennenwein. \textit{Experimental test
of the spin mixing interface conductivity concept}, Phys. Rev. Lett.
\textbf{111}, 176601 (2013).

\bibitem {Casanova}M. Isasa, A. Bedoya-Pinto, S. Velez, F. Golmar, F. Sanchez,
L. E. Hueso, J. Fontcuberta and F. Casanova. \textit{Spin Hall
magnetoresistance at Pt/CoFe$_{2}$O$_{4}$ interfaces and texture effects},
Appl. Phys. Lett. \textbf{105}, 142402 (2014).

\bibitem {Tokac}M. Toka\c{c}, S. A. Bunyaev, G. N. Kakazei, D. S. Schmool, D.
Atkinson, and A. T. Hindmarch. \textit{Interfacial structure dependent spin
mixing conductance in cobalt thin films }, Phys. Rev. Lett. \textbf{115},
056601 (2015).

\bibitem {BookMagnetismBlundell}S. Blundell, \textit{Magnetism in Condensed
Matter}, (Oxford University Press, New York, 2001).

\bibitem {BookMagnetismTMO}S. Maekawa, Y. Tohyama, S. E. Barnes, S. Ishihara, W. Koshibae, and G. Khaliullin. \textit{Physics of transition
metal oxides}, (Springer, Berlin, 2004).

\bibitem {BookSkomski1}R. Skomski, \textit{Simple Models of Magnetism},
(Oxford University Press, Croydon, 2008).

\bibitem {Ale}{A. B. Cahaya, A. O. Leon, M. Rahimi, and G. E. W. Bauer (unpublished)}

\bibitem {RudermanKittel}M.A. Ruderman and C. Kittel. \textit{Indirect
exchange coupling of nuclear magnetic moments by conduction electrons}, Phys.
Rev. \textbf{96}, 99 (1954).

\bibitem {Kasuya}T. Kasuya, Prog. Theor. \textit{A Theory of Metallic Ferro-
and Antiferromagnetism on Zener's Model} Phys. \textbf{16}, 45 (1956).

\bibitem {Yosida}K. Yosida. \textit{Spin Polarization of Conduction Electrons
Due to s-d Exchange Interaction}, Phys. Rev. \textbf{106}, 893 (1957).

\bibitem {Zhou}L. Zhou, J. Wiebe, S. Lounis, E. Vedmedenko, F. Meier, S.
Bl\"{u}gl, P. H. Dederichs and R. Wiesendanger. \textit{Strength and
directionality of surface Ruderman-Kittel-Kasuya-Yosida interaction mapped on
the atomic scale}, Nat. Phys. \textbf{6}, 187 (2010).

\bibitem {freeman}A. J. Freeman and R. E. Watson. \textit{Theoretical
investigation of some magnetic and spectroscopic properties of rare-earth
ions}, Phys. Rev. \textbf{127}, 2058 (1962).

\bibitem {waber}J. T. Waber and D. T. Cromer. \textit{Orbital Radii of Atoms
and Ions}, J. Chem. Phys. \textbf{42}, 4116 (1965).

\bibitem {JahnTeller}C. Housecroft, and A.G. Sharpe, \textit{Inorganic
Chemistry,} 3rd Ed., (Prentice Hall, London, 2008).

\bibitem {Auxetic}M. Valant, A.-K. Axelsson, F. Aguesse, and N. M. Alford.
\textit{Molecular Auxetic Behavior of Epitaxial Co-Ferrite Spinel Thin Film}, Adv. Funct. Mater. \textbf{20}, 644 (2010).

\bibitem {Auxetic2}M. N. Iliev, D. Mazumdar, J. X. Ma, A. Gupta, F. Rigato,
and J. Fontcuberta. \textit{Monitoring B-site ordering and strain relaxation
in NiFe$_{2}$O$_{4}$ epitaxial films by polarized Raman spectroscopy}, Phys.
Rev. B \textbf{83}, 014108 (2011).

\bibitem {Larsen}U. Larsen. \textit{A simple derivation of the sd exchange
interaction}, J. Phys. C: Solid St. Phys. \textbf{4}, 1835 (1971).

\bibitem {Davis}D. Davis. \textit{Thomas-Fermi screening in one dimension},
Phys. Rev. B \textbf{7}, 129 (1973).

\bibitem {Adachi}H. Adachi, K. Uchida, E. Saitoh and S. Maekawa.
\textit{Theory of the spin Seebeck effect}, Rep. Prog. Phys. \textbf{76},
036501 (2013).

\bibitem {Cahaya}A.B. Cahaya, O.A. Tretiakov and G.E.W. Bauer. \textit{Spin
Seebeck power conversion}, IEEE Trans. Magn. \textbf{51}, 0800414 (2015).

\bibitem {Lindhard}B. Mihaila, \textit{Lindhard function of a d-dimensional
Fermi gas} arXiv:1111.5337 [cond-mat.quant-gas].

\bibitem {Chen}Y.-T. Chen, S. Takahashi, H. Nakayama, M. Althammer, S. T. B.
Goennenwein, E. Saitoh, and G. E. W. Bauer. \textit{Theory of spin Hall
magnetoresistance}, Phys. Rev. B \textbf{87}, 144411 (2013).

\bibitem {Dieny}B. Dieny and M. Chshiev. \textit{Perpendicular magnetic
anisotropy at transition metal/oxide interfaces and applications}, Rev. Mod.
Phys. \textbf{89}, 025008 (2017).

\bibitem {SpinMixinConductance2005}M. Zwierzycki, Y. Tserkovnyak, P.J. Kelly,
A. Brataas, and G.E.W. Bauer. \textit{First-principles study of magnetization
relaxation enhancement and spin transfer in thin magnetic films}, Phys. Rev. B
\textbf{71}, 064420 (2005).

\bibitem {SPAFM}R. Cheng, J. Xiao, Q. Niu, and A. Brataas. \textit{Spin
Pumping and Spin-Transfer Torques in Antiferromagnets}. Phys. Rev. Lett.
\textbf{113}, 057601 (2014).

\bibitem {Takei}S. Takei, B. I. Halperin, A. Yacoby, and Y. Tserkovnyak.
\textit{Superfluid spin transport through antiferromagnetic insulators}. Phys.
Rev. B \textbf{90}, 094408 (2014).

\bibitem {StrainCFO}F. Rigato, J. Geshev, V. Skumryev, and J. Fontcuberta.
\textit{The magnetization of epitaxial nanometric CoFe$_{2}$O$_{4}$ (001)
layers}. J. Appl. Phys. \textbf{106}, 113924 (2009).

\bibitem {Dhakal}T. Dhakal, D. Mukherjee, R. Hyde, P. Mukherjee, M. H. Phan,
H. Srikanth, and S. Witanachchi. \textit{Magnetic anisotropy and field
switching in cobalt ferrite thin films deposited by pulsed laser ablation}, J.
Appl. Phys. \textbf{107}, 053914 (2010).

\bibitem {Takahashi}S. Takahashi and S. Maekawa. \textit{Spin current, spin
accumulation and spin Hall effect}, Sci. Technol. Adv. Mater. \textbf{9}
014105 (2008).


\end{thebibliography}
\end{document}